\documentclass[11pt]{article}
\usepackage{amsmath,amsthm,latexsym,amssymb,amsfonts,epsfig}

\newcommand{\Man}{\Sigma}

\DeclareMathOperator{\porder}{\mathcal{P}}   

\makeatletter
\@addtoreset{equation}{section}
\makeatother

\newtheorem{Theorem}{Theorem}[section]
\newtheorem{Definition}{Definition}[section]
\newtheorem{Lemma}{Lemma}[section]
\newtheorem{Corollary}{Corollary}[section]

\def\be{\begin{equation}}
\def\ee{\end{equation}}
\def\ba{\begin{eqnarray}}
\def\ea{\end{eqnarray}}
\def\a{{\cal A}}
\def\ab{\overline{\a}}

\def\Nl{{\mathchoice
{\setbox0=\hbox{$\displaystyle\rm N$}\hbox{\hbox to0pt
{\kern0.4\wd0\vrule height0.9\ht0\hss}\box0}}
{\setbox0=\hbox{$\textstyle\rm N$}\hbox{\hbox to0pt
{\kern0.4\wd0\vrule height0.9\ht0\hss}\box0}}
{\setbox0=\hbox{$\scriptstyle\rm N$}\hbox{\hbox to0pt
{\kern0.4\wd0\vrule height0.9\ht0\hss}\box0}}
{\setbox0=\hbox{$\scriptscriptstyle\rm N$}\hbox{\hbox to0pt
{\kern0.4\wd0\vrule height0.9\ht0\hss}\box0}}}}
\def\Zl{{\mathchoice
{\setbox0=\hbox{$\displaystyle\rm Z$}\hbox{\hbox to0pt
{\kern0.4\wd0\vrule height0.9\ht0\hss}\box0}}
{\setbox0=\hbox{$\textstyle\rm Z$}\hbox{\hbox to0pt
{\kern0.4\wd0\vrule height0.9\ht0\hss}\box0}}
{\setbox0=\hbox{$\scriptstyle\rm Z$}\hbox{\hbox to0pt
{\kern0.4\wd0\vrule height0.9\ht0\hss}\box0}}
{\setbox0=\hbox{$\scriptscriptstyle\rm Z$}\hbox{\hbox to0pt
{\kern0.4\wd0\vrule height0.9\ht0\hss}\box0}}}}
\def\Ql{{\mathchoice
{\setbox0=\hbox{$\displaystyle\rm Q$}\hbox{\hbox to0pt
{\kern0.4\wd0\vrule height0.9\ht0\hss}\box0}}
{\setbox0=\hbox{$\textstyle\rm Q$}\hbox{\hbox to0pt
{\kern0.4\wd0\vrule height0.9\ht0\hss}\box0}}
{\setbox0=\hbox{$\scriptstyle\rm Q$}\hbox{\hbox to0pt
{\kern0.4\wd0\vrule height0.9\ht0\hss}\box0}}
{\setbox0=\hbox{$\scriptscriptstyle\rm Q$}\hbox{\hbox to0pt
{\kern0.4\wd0\vrule height0.9\ht0\hss}\box0}}}}
\def\Rl{{\mathchoice
{\setbox0=\hbox{$\displaystyle\rm R$}\hbox{\hbox to0pt
{\kern0.4\wd0\vrule height0.9\ht0\hss}\box0}}
{\setbox0=\hbox{$\textstyle\rm R$}\hbox{\hbox to0pt
{\kern0.4\wd0\vrule height0.9\ht0\hss}\box0}}
{\setbox0=\hbox{$\scriptstyle\rm R$}\hbox{\hbox to0pt
{\kern0.4\wd0\vrule height0.9\ht0\hss}\box0}}
{\setbox0=\hbox{$\scriptscriptstyle\rm R$}\hbox{\hbox to0pt
{\kern0.4\wd0\vrule height0.9\ht0\hss}\box0}}}}
\def\Cl{{\mathchoice
{\setbox0=\hbox{$\displaystyle\rm C$}\hbox{\hbox to0pt
{\kern0.4\wd0\vrule height0.9\ht0\hss}\box0}}
{\setbox0=\hbox{$\textstyle\rm C$}\hbox{\hbox to0pt
{\kern0.4\wd0\vrule height0.9\ht0\hss}\box0}}
{\setbox0=\hbox{$\scriptstyle\rm C$}\hbox{\hbox to0pt
{\kern0.4\wd0\vrule height0.9\ht0\hss}\box0}}
{\setbox0=\hbox{$\scriptscriptstyle\rm C$}\hbox{\hbox to0pt
{\kern0.4\wd0\vrule height0.9\ht0\hss}\box0}}}}
\def\Hl{{\mathchoice
{\setbox0=\hbox{$\displaystyle\rm H$}\hbox{\hbox to0pt
{\kern0.4\wd0\vrule height0.9\ht0\hss}\box0}}
{\setbox0=\hbox{$\textstyle\rm H$}\hbox{\hbox to0pt
{\kern0.4\wd0\vrule height0.9\ht0\hss}\box0}}
{\setbox0=\hbox{$\scriptstyle\rm H$}\hbox{\hbox to0pt
{\kern0.4\wd0\vrule height0.9\ht0\hss}\box0}}
{\setbox0=\hbox{$\scriptscriptstyle\rm H$}\hbox{\hbox to0pt
{\kern0.4\wd0\vrule height0.9\ht0\hss}\box0}}}}
\def\Ol{{\mathchoice
{\setbox0=\hbox{$\displaystyle\rm O$}\hbox{\hbox to0pt
{\kern0.4\wd0\vrule height0.9\ht0\hss}\box0}}
{\setbox0=\hbox{$\textstyle\rm O$}\hbox{\hbox to0pt
{\kern0.4\wd0\vrule height0.9\ht0\hss}\box0}}
{\setbox0=\hbox{$\scriptstyle\rm O$}\hbox{\hbox to0pt
{\kern0.4\wd0\vrule height0.9\ht0\hss}\box0}}
{\setbox0=\hbox{$\scriptscriptstyle\rm O$}\hbox{\hbox to0pt
{\kern0.4\wd0\vrule height0.9\ht0\hss}\box0}}}}

\title{On the Superselection Theory of the Weyl Algebra for Diffeomorphism
Invariant Quantum Gauge Theories}
\author{
Hanno Sahlmann\thanks{hanno@gravity.psu.edu},\\
Center for Gravitational Physics and Geometry, \\
The Pennsylvania State University, University Park, PA, USA\\
\\
Thomas Thiemann\thanks{thiemann@aei-potsdam.mpg.de,
tthiemann@perimeterinstitute.ca}\thanks{New Address:
The Perimeter Institute for Theoretical Physics
and Waterloo University, Waterloo, Ontario, Canada},\\
MPI f\"ur Gravitationsphysik, Albert-Einstein-Institut, \\
Am M\"uhlenberg 1, 14476 Golm near Potsdam, Germany
}
\date{{\small Preprint AEI-2003-025, PI-2003-001, CGPG-03/2-1, ESI-1284}}
\begin{document}
\maketitle
\begin{abstract}
Much of the work in loop quantum gravity and quantum geometry rests on a 
mathematically rigorous integration theory on spaces of distributional
connections. Most notably, a diffeomorphism invariant representation
of the algebra of basic observables of the theory, the
Ashtekar-Lewandowski representation, has been
constructed. This representation is singled out by its mathematical
elegance, and up to now, no other diffeomorphism invariant
representation has been constructed. 
This raises the question whether it is unique in a precise sense. 

In the present article we take steps towards answering this
question. Our main result is that upon imposing relatively mild
additional assumptions, the AL-representation is indeed unique. 
As an important tool which is also interesting in its own right,
we introduce a $C^*$-algebra which is very similar to the
Weyl algebra used in the canonical quantization of free quantum field
theories. 
\end{abstract}

\section{Introduction}
\label{s1}
Canonical, background independent quantum field theories of connections 
\cite{1} play a fundamental role in the program of canonical 
quantization of general relativity (including all types of matter),
sometimes called loop quantum gravity or quantum general relativity.
For a review geared to mathematical physicists see \cite{2}, for a
general overview \cite{h1}). 

The classical canonical theory can be formulated in terms of smooth 
connections $A$ on principal $G-$bundles over a $D-$dimensional spatial 
manifold $\Man$ for a compact gauge group $G$ and smooth sections of an 
associated (under the adjoint representation) vector bundle of 
Lie$(G)-$valued vector densities $E$ of weight one. The pair 
$(A,E)$ coordinatizes an infinite dimensional symplectic manifold 
$({\cal M},\Man)$ whose (strong) symplectic structure $s$ is such
that $A$ and $E$ are canonically conjugate.

In order to quantize $({\cal M},s)$, it is necessary to smear the 
fields $A,E$. This has to be done in such a way, that the smearing 
interacts well with two fundamental automorphisms of the principal 
$G-$bundle, namely the vertical automorphisms formed by $G-$gauge 
transformations and the horizontal automorphisms formed by Diff$(\Man)$ 
diffeomorphisms. These requirements naturally
lead to holonomies and electric fluxes, that is, exponentiated 
(path-ordered) 
smearings of the connection over $1-$dimensional submanifolds $e$ of $\Man$ 
as well as smearings of the electric field over $(D-1)-$dimensional 
submanifolds $S$
\begin{equation*}
  h_e[A]=\porder \exp \int_e A,\qquad E_{S,f}[E]=\int_S *E_I f^I
\end{equation*}
These functions on $\cal M$ generate a closed Poisson$^\ast-$algebra 
$\cal P$ and separate the points of $\cal M$. They do not depend on a
choice of coordinates nor on a background metric. Therefore, 
diffeomorphisms and gauge transformations act on these variables in a
remarkably simple way: Let $\varphi$ be a diffeomorphism of $\Man$, then 
\begin{equation*}
  \alpha_\varphi(h_e)=h_{\varphi^{-1} e},\qquad
  \alpha_\varphi(E_{S,f})=E_{\varphi^{-1} S,\varphi^*f}.
\end{equation*}
Similarly let $g:\Man\rightarrow G$ be a gauge transformation, then 
\begin{equation*}
\alpha_g(h_e)=g(a)h_{e}g^{-1}(b),\qquad \alpha_g(E_{S,f})=E_{
  S,g^{-1}fg}
\end{equation*}
where $a$ is the starting point of $e$ and $b$ the endpoint. 
 
Quantization now means to promote $\cal P$ to an abstract $^\ast-$algebra
$\mathfrak{A}$ and to look for its representations. However, for
physical reasons we are not interested in arbitrary representations
but those fulfilling the following criteria:
\begin{itemize}
\item[i)] {\it Irreducibility}\\
The representation space ${\cal H}_\pi$ should contain no proper invariant
subspaces, i.e. the span of vectors $\pi(a)v$ should be dense 
in ${\cal H}_pi$ for any vector $v\in {\cal H}$.

Irreducible representations are the building blocks of the
representation theory. If their structure is clarified, more general
representations can be constructed from and analyzed in terms of them.  
\item[ii)] {\it Diffeomorphism  and Gauge Invariance}\\
Diffeomorphism and gauge transformations are fundamental symmetries
of the theory, so if we do not consider a scenario of spontaneous 
symmetry breaking, they should be symmetries of the ground state of
the quantum theory as well. 

Thus in our setting we require that there is a at least one symmetric 
state $\Omega_\pi$ in the representation space. More precisely, for
the expectation value 
$\omega_\pi(.):=<\Omega_\pi,.\Omega_\pi>_{{\cal H}_\pi}$ in that
state, we require invariance: 
\begin{equation*}
  \omega_\pi\circ\alpha_\varphi=\omega_\pi,\qquad 
\omega_\pi\circ\alpha_g=\omega_\pi
\end{equation*}
for all diffeomorphisms $\varphi$ and gauge transformations $g$.  
\end{itemize}
It is remarkable that so far only one representation has been found 
which satisfies our assumptions: This is the Ashtekar -- Isham -- 
Lewandowski representation $\pi_0$ on a Hilbert space 
${\cal H}_0=L_2(\ab,d\mu_0)$ where $\ab$ is the Ashtekar -- Isham
space of distributional connections (the spectrum of a certain Abelian
C$^\ast-$algebra) and $\mu_0$ is the Ashtekar -- Lewandowski measure.
Historically, first Ashtekar and Isham \cite{6} were looking for a 
natural distributional extension $\ab$ of the space $\a$ of smooth 
connections, which could serve as the support for gauge invariant measures.  
Then Ashtekar and Lewandowski found a natural, cylindrical measure 
\cite{7} which was shown to have a unique $\sigma-$additive extension 
$\mu_0$ by Marolf and Mour\~ao \cite{8}. This measure turned out to be 
diffeomorphism invariant. More general diffeomorphism invariant measures 
were found by Baez \cite{9}, however, in contrast to $\mu_0$ they are not 
faithful.
That the resulting Hilbert space ${\cal H}_0$ indeed carries a 
representation of the holonomy -- flux algebra was shown only later 
in \cite{1}, essentially that representation $\pi_0$ results by having 
connections and electric fields respectively act as multiplication and 
functional derivative operators respectively.

The present work was inspired by the question whether the fact that ${\cal 
H}_0$ is the only representation found so far which satisfies our 
criteria in fact means that it is the unique representation. 
In this article we show that upon imposing two additional and
rather technical conditions on the representations, the question can be
answered affirmatively: Under these assumptions, the
Ashtekar-Lewandowski representation is indeed unique.  

Work towards settling this questions has begun in \cite{10}, however
the results obtained there rest on assumptions that exclude the
interesting cases, most notably that of a noncommutative gauge group.  
However it might still be interesting for the reader to take a look
at \cite{10} since the discussion there is much less burdened by the
technical subtleties that arise in the general case. 

During the completion of this article, a very interesting work has been
published by Okolow and Lewandowski \cite{25} that aims at settling
the very same question raised in this article. Their method of proof
and in part also their assumptions differ from the ones used in the
present article, so it is very instructive to compare the two
approaches. The hope is that combining methods of the present
paper with those of \cite{25} enables one to prove a completely 
general and satisfactory uniqueness theorem.  

Before we conclude this introduction, let us discuss the
subtleties that arise due to our general setting as well as the
additional assumptions we are going to make. 

The first problem that arises comes from the fact that the flux
operators are unbounded and so one has to worry about domain
problems. In our approach, we will try to circumvent these problems by
not working with the fluxes directly but with their exponentiated
counterparts. 
More precisely, we 
will consider the abstract Weyl algebra formed from holonomies and 
exponentiated electric fluxes and represent them as bounded operators 
on a Hilbert space . This algebra can be equipped with a 
$C^\ast-$norm so that $\mathfrak{A}$ turns into a $C^\ast-$algebra and we 
therefore have the powerful representation theory of $C^\ast-$algebras 
at our disposal. However, we will require that the representations
under considerations will be weakly continuous for the unitary groups
generated by exponentiated fluxes. Therefore their selfadjoint
generators, the fluxes themselves, will be well defined operators.
In the case of an Abelian gauge group, this approach enables us to
completely circumvent any specification of the domains of the
fluxes. Due to technical complications for non-Abelian gauge groups,
we will however have to make such a specification in that case. This
is the first of the two requirements in addition to i) and ii) above
that we make in order to prove our uniqueness result.    

It is interesting to note that, at least for the case of an Abelian
gauge group, our theorem could be compared to von Neumann's theorem \cite{11}
(uniqueness of weakly continuous, irreducible representations of the 
Weyl C$^\ast-$algebra of the phase space $({\cal M}=\mathbb{R}^{2N},
\sigma=\sum_{a=1}^N dp_a\wedge dq^a)$ with $N<\infty$ up to unitary 
equivalence) since it also makes use of irreducibility 
and continuity. The surprise is that our theorem
holds for an infinite number of degrees of freedom and that continuity 
is required only for one half of the variables (in fact, connections only
form an affine space and not a vector space, so continuity of holonomies 
is even hard to formulate) while in background dependent quantum field 
theories we are faced with an uncountably infinite number of unitarily 
inequivalent representations of the canonical commutation relations
\cite{12}. There, a unique representation is usually selected by using
Lorentz invariance and a specific dynamics, in that sense it is a 
{\it dynamical uniqueness}. However, while we use spatial 
diffeomorphism invariance, in our case we do not make use of any 
particular dynamics such as the Hamiltonian constraint of quantum general
relativity \cite{13} and in that sense it is a {\it kinematical 
uniqueness}.

This comparison leads us to the second subtlety and the corresponding
additional assumption: The requirements i) and ii) guarantee that the 
the action of the automorphisms on algebra elements can be unitarily
implemented in the representation. However there is a priory little
control about the details of the action of these unitary operators in
the Hilbert space. We will point out that there is a ``natural'' way for
them to act in the representation Hilbert space, and we will require
that this natural action is realized in the representations we
consider. This is, however, a priori not the most general
possibility. Two scenarios can be envisioned: In the first, one can
actually show that the natural action is in fact the only possible
one, and then our uniqueness result would be general.
In the second, there are actually other viable unitary actions of the
diffeomorphisms, and this in turn might lead to a classifications of
the representations studied in terms of unitary representations of the
diffeomorphism group. The picture then would be very similar to that
obtained in the case of free quantum field theories, where Poincare
invariant representations can be classified by unitary representations
of the Poincare group. Both scenarios would be very interesting in
their own ways, and we happily await a future settling of this
question.  

To summarize, a completely general and satisfactory picture of the
diffeomorphism and gauge invariant representations of the algebra of
holonomies and fluxes has not yet emerged. However, the results of the
present work and that of \cite{25,10} point to the fact that 
diffeomorphism invariance is an extremely strong requirement and could
mean that in background independent quantum field theories there 
is much less quantization freedom than in background dependent ones.
\\[1ex]
To finish, let us give an overview of the structure of the rest of
present work:\\
\\
In section \ref{s2} we recall from \cite{1} the essentials of the 
classical formulation 
of canonical, background independent theories of connections, that is,
the symplectic manifold $({\cal M},\sigma)$ and the corresponding 
classical Poisson$^\ast-$algebra $\cal P$ generated by holonomies and 
electric fluxes.

In section \ref{s3} we define the abstract $^\ast-$algebra 
$\mathfrak{A}$ and recall from \cite{10} the general representation theory 
of $\mathfrak{A}$.

In section \ref{s4} we implement irreducibility and spatial diffeomorphism 
invariance and prove our uniqueness theorem.

Finally, in appendix \ref{sa} we present a simpler proof of our main result 
using an additional, natural assumption. We also report some additional
technical information in appendix \ref{sb}.

\section{Preliminaries}
\label{s2}

Let $\Man$ be an analytic, connected and orientable $D-$dimensional
manifold and $G$ a compact, connected  gauge group. A principal 
$G-$bundle $P$ over $\Man$ is determined by its local trivializations 
$\phi_I:\;U_I\times G\to P$ subordinate to 
an atlas $\{U_I\}$ of $\Man$. These give rise to local, smooth 
$G-$valued 
functions $g_{IJ}:\;U_I\cap U_J\to G$ on $\Man$, called transition 
function cocycles. A connection over $P$ can be thought of as a 
collection $\{A_I\}$ of smooth, Lie$(G)-$valued one-forms over the 
respective charts $U_I$ subject to the gauge covariance condition 
$A_I=-dg_{IJ} g_{IJ}^{-1}+\mbox{Ad}_{g_{IJ}}(A_J)$ over $U_I\cap U_J$.
The space of smooth connections ${\a}$ over $P$ therefore depends on the 
bundle $P$ but we will abuse notation in not displaying this dependence.

Similarly, we define a vector bundle $E_P$ associated to $P$ under the 
adjoint representation whose typical fiber is a Lie$(G)-$valued 
$(D-1)-$form on $P$. An electric field is a local section of $E_P$ which 
we may think of as a collection $\{E_I\}$ of Lie$(G)-$valued 
$(D-1)-$forms on $\Man$ subject to the gauge covariance condition 
$E_I=\mbox{Ad}_{g_{IJ}}(E_J)$ over $U_I\cap U_J$. The space of 
smooth electric fields $\cal E$ over $P$ depends on $P$ as well but 
the dependence is also not displayed.

The space $\a$ can be given the structure of a manifold 
modeled on a Banach space in the usual way (see e.g. \cite{14,1}).
Consider now the cotangent bundle ${\cal M}:=T^\ast(\a)$. Since 
$\a$ is a Banach manifold, also $\cal M$ is and, moreover, we may
identify $\cal E$ with the sections of $\cal M$ together with the 
induced topology. The cotangent bundle $\cal M=\a\times {\cal E}$ can be 
equipped with the following (strong, see e.g. \cite{15}) symplectic 
structure: Let $\mbox{Tr}:\;\mbox{Lie}(G)\times\mbox{Lie}(G)\mapsto 
\mathbb{C}$ be a natural $Ad_G-$invariant metric on Lie$(G)$ then 
there is a natural pairing ${\cal E}\times \a\to \mathbb{C}$ defined by
\be \label{2.1}
(F,f)\mapsto F(f):=\int_\Man \mbox{Tr}(F\wedge f)
\ee
Since in our case the section of the tangential space of $\cal M$ can be 
identified with points in $\cal M$, we may define the symplectic 
structure by
\be \label{2.2}
s:\;T({\cal M})\times T({\cal M})\to \mathbb{C};\;
((F,f),(F',f'))\mapsto F(f')-F'(f)
\ee
In a concrete gauge field theory the right hand side will be multiplied 
by a constant which depends on the coupling constant of the theory.
In order not to clutter our formulae we will assume that $A$ and $E$ 
respectively have dimension cm$^{-1}$ and cm$^{-(D-1)}$ respectively and 
we set $\hbar=1$ for simplicity.

The Poisson bracket is uniquely defined by 
\be \label{2.3}
\{.,.\}:\;C^\infty({\cal M})\times C^\infty({\cal M});\;
(a,b)\mapsto \chi_a(b)
\ee
where the Hamiltonian vector field $\chi_a$ on $\cal M$ defined by
$a\in C^\infty({\cal M})$ is uniquely defined by $i_{\chi_a}s+da=0$.
It is easy to see that the usual Poisson bracket $^\ast-algebra$ 
${\cal P}':=C^\infty({\cal M})$ is generated
from the basic canonical bracket relations
\be \label{2.4}
\{F(A),F'(A)\}=\{E(f),E(f')\}=0,\;\;\{E(f),F(A)\}=F(f)
\ee
and the reality conditions
\be \label{2.5}
\overline{E(f)}=E(\bar{f}),\;
\overline{F(A)}=\bar{F}(A)
\ee
The relations (\ref{2.4}) display $A,E$ respectively as the canonically
conjugate configuration and momentum degrees of freedom and therefore 
$\a$ is called the classical configuration space.

The algebra ${\cal P}'$ is, however, not what we are interested in for 
several reasons:
\begin{itemize}
\item[i)] {\it Gauge Invariance}\\
The objects $F(A),E(f)$ depend heavily on our choice of trivialization of 
$P$. It will be very hard to construct gauge invariant quantities from
them, in which we are ultimately interested. In order to do that, we must
work with basic functions on $\cal M$ which are different from the 
canonical functions $F(A),E(f)$. Of course, these problems could be 
avoided by fixing a gauge, however, there is no canonical gauge and 
most gauges are plagued by the Gribov problem.
\item[ii)] {\it Background Independence}\\
Even when ignoring the just mentioned problems, it is rather hard 
to construct spatially diffeomorphism invariant (background 
independent) representations of ${\cal P}'$, in fact, to the best 
of our knowledge such representations have not been constructed.
To see where the problem is, suppose that we want to construct
a representation of the form ${\cal H}=L_2({\cal S}',d\mu)$ where 
${\cal S}'$ is the space of tempered distributions on $\Man$
(that is, the topological dual of the space $\cal S$ of functions
of rapid decrease) 
and $\mu$ is a measure thereon. This is the form of the representation 
for free field theories \cite{16}. Notice that the 
nuclear topology on $\cal S$ does not refer to any background structure
except for the differentiable structure of $\Man$,
so there is no problem up to this point. The problem arises when 
we define the measure $\mu$ via its generating functional
$\mu(F):=\mu(\exp(iF(.)))$. For instance, if $\mu$ is a 
(generalized) free (Gaussian) measure,
then $\mu(F)=\exp(-F(C\cdot F)/2)$ where $C$ is a {\it background metric
dependent} appropriate covariance which is needed in order to contract
indices in the appropriate way. Interacting measures in more than 
three spacetime dimensions have not been constructed so far.
\end{itemize}
A solution to the first problem was suggested for canonical quantum 
Yang-Mills theories already by Gambini et. al. \cite{17} and for loop 
quantum gravity by Jacobson, Rovelli and Smolin \cite{18}. 
The idea is to work with holonomies and electric fluxes. We will 
explain in detail what we mean by that, because it will be important 
for what follows. For more details, see \cite{2}.
\begin{Definition} \label{def2.1} ~~~\\
i) $\cal C$ is the set of piecewise analytic, continuous, oriented, 
compactly supported, parameterized curves embedded in $\Man$.
We denote by $b(c),f(c)$ the beginning and final point of $c$ and consider
the range $r(c)$ as the image of the compact interval $[0,1]$ under $c$. 
\\
ii) If $b(c_2)=f(c_1)$ we define composition 
$(c_1\circ c_2)(t)=c_1(2t)$ if $t\in[0,\frac{1}{2}]$ and  
$(c_1\circ c_2)(t)=c_2(2t-1)$ if $t\in[\frac{1}{2},1]$. Inversion is 
defined by $c^{-1}(t):=c(1-t)$.
\\
iii) We call $c,c'\in {\cal C}$ equivalent, $c\sim c'$, iff $c,c'$
differ by a finite number of reparameterizations and retracings
(a segment of a curve of the form $s^{-1}\circ s'$). The set of 
equivalence classes $p$ in $\cal C$ is denoted as the set of paths 
$\cal Q$. The functions $b,f$ and the operations $\circ,^{-1}$ extend from 
$\cal C$ to $\cal Q$.
\\
iv) An edge $e\in {\cal Q}$ is a path for which an entire analytic
representative $c_e\in {\cal C}$ exists. For edges the function 
$r$ extends as $r(e):=r(c_e)$.
\\
v) An oriented graph $\gamma$ is determined by a finite number of edges 
$e\in E(\gamma)$ which intersect at most in their boundaries, called 
the vertex set $V(\gamma)$.
\end{Definition}
It is important to realize that in contrast to $\cal C$ the set $\cal Q$
is a groupoid with objects the points $x\in \Man$ and with the sets 
of morphisms given by Mor$(x,y)=\{p\in {\cal Q};\;b(p)=x,\;f(p)=y\}$.
The notion of paths is motivated by the algebraic properties of the 
holonomy.
\begin{Definition} \label{def2.2} ~~~~~\\
For $A\in \a$ and $p\in {\cal Q}$ we define $A(p):=h_{A,p}(1)$ where
$h_{A,p}:\;[0,1]\to G$ is uniquely defined by the parallel 
transport equation
\be \label{2.6}       
\frac{d}{dt} h_{A,p}(t)=h_{A,p}(t)A_a(c_p(t)) \dot{c}^a_p(t),\;
h_{A,p}(0)=1_G
\ee
if $p$ is in the domain of a chart, where $a=1,..,D$ denote tensorial 
indices and $[c_p]=p$ is a representative. 
\end{Definition}
Due to the covariance condition on connections under change of local 
trivialization one can show that (\ref{2.6}) can be extended 
unambiguously (up to a gauge transformation) to the case that $p$ is not 
within the domain of a 
chart. This follows from the fact that it comes from the horizontal lift 
of $c_p$ which is globally defined. The virtue of definition (\ref{2.6})
is that it displays $A\in \a$ as an a groupoid morphism 
$A\in \ab:=\mbox{Hom}({\cal Q},G)$. In fact, $A(c)=A([c])$ is 
reparameterization invariant and $A(p_1\circ p_2)=A(p_1)A(p_2),\;
A(p^{-1})=A(p)^{-1}$. Under gauge transformations $g\in 
\mbox{Fun}(\Man,G)$ we find $A^g(p)=g(b(p))A(p)g(f(p))^{-1}$ which
implies that e.g. traces of holonomies along closed paths are gauge 
invariant. Thus, it is relatively easy to construct gauge invariant 
functions of the connection from holonomies !
 
The worry is of course, that $A(p)$ is smeared only in one dimension
rather than three such as $F(A)$ was. In order to still obtain a 
well-defined Poisson algebra, the electric field therefore must be 
smeared in at least $D-1$ dimensions. This can be done as follows:\\
Let $S$ be an open, connected, simply connected, analytic, oriented, 
compactly supported  
$(D-1)-$dimensional submanifold of $\Man$, called a surface in what 
follows, let $x_0\in S$ and for $x\in 
S$ let $c_{x_0,x}\in 
{\cal C}$ with $b(c_{x_0,x})=x_0,\;f(c_{x_0,x})=x,\;r(c_{x_0,x})\subset 
S$. Then we define 
\be \label{2.7}
E'(S):=\int_S \mbox{Ad}_{A(p_{x_0,x})}(E(x))
\ee
It is easy to see that under gauge transformations 
$E^{\prime g}(S)=\mbox{Ad}_{g(x_0)}(E'(S))$ so that for instance 
$\mbox{Tr}(E'(S)^2)$
is gauge invariant. Notice that the holonomies involved in (\ref{2.7})
are only necessary if $G$ is non-Abelian. The ugly feature of $E'(S)$
is that it depends not only on $S$ but also on $x_0,p_{x_0,x}$. 
Consider therefore the non-covariant object 
\be \label{2.8}
E_n(S):=\int_S E^j n_j
\ee
where $n_j,\;j=1,..,\dim(G)$ is a Lie$(G)-$valued scalar. It  
will be sufficient to normalize the components corresponding to the non 
Abelian generators by $\delta^{jk} n_j(x) n_k(x)=1$
because we need it only in order to allow for local gauge transformations
which have the effect of simply rotating $n_j(x)$ locally.
For the same reason we restrict to $n_j(x)=const.$ if $j$ corresponds to 
an Abelian generator.
The idea is that while (\ref{2.8}) does not transform simply, we can 
still construct gauge invariant functions from it using a limiting 
procedure that involves making the surfaces smaller and smaller while 
making their number larger and larger at the same time. Examples are 
provided by the length, area and volume functionals already mentioned.
Thus, the $E_n(S)$ serve as an intermediate objects to build more 
complicated but gauge invariant composite objects and this is why
we want them to be represented as well-defined operators later on, because 
once they are defined, the composite operators can be defined as well.

For the purposes of this paper we will make also the following additional
technical assumption: Notice that if $S=S_1\cup S_2$ is the disjoint 
union of surfaces then we have $E_n(S)=E_n(S_1)+E_n(S_2)$. Thus we know 
the flux $E_n(S)$ if we know it for every connected surface $S$. If 
$S$ is a connected surface we can triangulate it into $(D-1)-$simplices 
$\Delta$ and we have $E_n(S)=\sum_\Delta E_n(\Delta)$ even if the different
$\Delta$ overlap in faces, since they are of measure zero. Now each 
$(D-1)-$simplex can be decomposed into $D$, $(D-1)-$dimensional, cubes
by choosing an interior point of $\Delta$, connecting it with an interior 
point
of each of its boundary $(D-2)-$simplices, connecting those points with 
an interior point of each of its boundary $(D-3)-$simplices etc. 
Thus, we know each $E_n(S)$ if we know it for each $E_n(\Box)$ where 
$\Box$ is a $(D-1)-$cube. The assumption that we now make is the following:
We choose precisely one $(D-2)-$face of $\Box$ open while all others are 
closed. In other words, if $\overline{\Box}$ denotes the closure of $\Box$
and $\bar{F}$ the closure of one of its faces $F$ then 
$\Box=\overline{\Box}-\overline{F}$. The classical fluxes satisfy 
$E_n(\Box)=E_n(\overline{\Box})$ so this seems to be an innocent assumption.
However, it will turn out to be crucial in the quantum theory. From now on
we allow only compactly supported, analytical, oriented 
surfaces $S$ which can be written as a disjoint union $S=\cup_\Box \Box$ 
of such cubes $\Box$ with the specified boundary properties. Since the 
classical flux $E_n(S)$ through any $S$ can be written as a limit of  
fluxes through those special $S$, there is no loss of generality on the 
classical side. The decisive feature of such a cube $\Box$ 
is that we can choose a closed $(D-2)-$surface $S$ such that 
$\Box=\Box_1\cup\Box_2$ is a disjoint union with $S\subset\Box_1,\;
S\cap \Box_2=\emptyset,\;\overline{\Box_1} \cap \overline{\Box_2}=S$ and 
all three $\Box,\Box_1,\Box_2$ are analytically diffeomorphic. The reason 
for why this is important will become obvious only in section \ref{s4}.
In order that this works, we must restrict to $D\ge 2$ in what follows.
We feel that this assumption is not crucial for our result to hold, however,
it avoids tedious case by case considerations of the intersection 
structure of surfaces.
Thus, we ask whether the functions $A(p),E_n(S)$ generate a well-defined 
Poisson algebra which is induced from (\ref{2.4}). The answer is as 
follows \cite{19}:
\begin{Definition} \label{def2.3} ~~~~\\
i)\\
Given a graph $\gamma$ we define $p_\gamma:\;\a\to G^{|E(\gamma)|};\;
A\mapsto \{A(e)\}_{e\in E(\gamma)}$. A function $f$ is said to be 
cylindrical over $\gamma$ iff there exists a function $f_\gamma:\;
G^{|E(\gamma)|}\to \mathbb{C}$ such that $f=f_\gamma\circ p_\gamma$.
The functions cylindrical over $\gamma$ are denoted by Cyl$_\gamma$ and 
the $^\ast-$algebra of cylindrical functions is defined by 
Cyl$:=\cup_{\gamma\in \Gamma} \mbox{Cyl}_\gamma$ where $\Gamma$ is the
set of all compactly supported, oriented, piecewise analytic graphs.
Notice that $f\in \mbox{Cyl}_\gamma$ implies $f\in \mbox{Cyl}_{\gamma'}$ 
for any $\gamma\subset\gamma'$ and we identify the corresponding 
representatives.\\
ii)\\
The subalgebras Cyl$^n,\;n=0,1,2,..,\infty$ of Cyl consist of functions 
of the form $f=f_\gamma\circ p_\gamma$ where $f_\gamma\in 
C^n(G^{|E(\gamma)|})$.\\
iii)\\
Vector fields on $\a$ are defined as maps 
$Y:\;\mbox{Cyl}^n\to \mbox{Cyl}^{n-1}$ which satisfy the Leibniz rule and 
annihilate constants. We will denote them by Vec. \\
iv)\\
Given an open, compactly supported, connected, simply 
connected, oriented, analytic surface $S$ and a cylindrical function 
$f$ we can always find a graph $\gamma$ over which it is cylindrical and 
which is adapted to 
$S$ in the following sense: Any $e\in E(\gamma)$ belongs to precisely one 
of the following subsets $E_\ast(\gamma)$ of $E(\gamma)$ where\\
$E_{out}(\gamma)=\{e\in E(\gamma);\;e\cap S=\emptyset\}$,\\    
$E_{in}(\gamma)=\{e\in E(\gamma);\;e\cap \bar{S}=e\}$, \\   
$E_{up}(\gamma)=\{e\in E(\gamma);\;e\cap S=b(e),\;e$ points into the 
direction of $S\}$ and\\    
$E_{down}(\gamma)=\{e\in E(\gamma);\;e\cap S=b(e),\;e$ points into the 
opposite direction of $S\}$.\\
For $e\in E(\gamma)$ we define $\sigma(S,e):=0$ if 
$e\in E_{out}(\gamma)\cup E_{in}(\gamma)$ and we define
a) $\sigma(S,e)=1$ if $e\in E_{up}(\gamma)$ 
b) $\sigma(S,e)=-1$ if $e\in E_{down}(\gamma)$.\\
We have supplemented the regularization of the 
flux vector field, so far only discussed for open surfaces in the 
literature, to the case that $b(e)$ is a boundary point. Our condition 
is compatible with the additivity of fluxes. 
We can now define a real-valued vector field 
$Y_n(S)$ on Cyl by ($f=p_\gamma^\ast f_\gamma$)
\be \label{2.9}
Y_n(S) f:=p_\gamma^\ast (Y_n(S))_\gamma f_\gamma:=
p_\gamma^\ast(\sum_{e\in E(\gamma)} \sigma(e,S) n_j(b(e))
R^j_e f_\gamma 
\ee
where $R^j_e=R^j(h_e),
[p_\gamma^\ast h_e](A)=A(e)$ and 
$R^j=(\frac{d}{dt})_{t=0} L_{\exp(t\tau_j)}$
denotes the generator of  
left translations on $G$ and $(\tau_j)_{j=1}^{\dim(G)}$
is a basis of Lie$(G)$. One can check that the family of vector fields 
$\{(Y_n(S))_\gamma\}_{\gamma\in \Gamma}$ is indeed consistent, that is,
if $\gamma\subset\gamma'$ and $p_{\gamma'\gamma}:=p_\gamma\circ 
p_{\gamma'}^{-1}$ then for $f\in \mbox{Cyl}_\gamma$ we have 
$p_{\gamma'\gamma}^\ast [(Y^j_S)_\gamma f_\gamma]=
(Y^j_S)_{\gamma'} p_{\gamma'\gamma}^\ast f_\gamma$.\\
v)\\
Consider the Lie$^\ast-$algebra $V:=$Cyl$^\infty \times$Vec defined by 
\be \label{2.10}
[(f,Y),(f',Y')]:=(Y\cdot f'-Y'\cdot f,[Y,Y'])
\ee
where the $^\ast-$operation is just complex conjugation and 
where the second entry on the right hand side of (\ref{2.10}) denotes 
the Lie bracket of vector fields. We identify $f$ with $(f,0)$ and 
$Y$ with $(0,Y)$. The algebra $\cal P$ is defined as the free tensor
algebra over $V$ modulo the two -- sided ideal generated 
by elements of the form $u\otimes v-v\otimes u-[u,v]$ for any 
$u,v\in V$ (also called the universal enveloping algebra of 
$V$)\footnote{If there are any additional algebraic relations in 
$\cal P$ then we enlarge the ideal correspondingly.}.
In what follows we drop the tensor product symbol $\otimes$ as usual.
\end{Definition}
This definition answers our question in the following sense: Notice that 
since $A(e),E_n(S)$ are not smeared in $D$ dimensions, the Poisson bracket
(\ref{2.4}) is actually ill-defined. However, one can regularize 
these functions by fattening out $e,S$ to $D-$dimensional tubes $e_r$
and disks $S_r$ respectively where $r$ is some regularization parameter, 
compute the Poisson brackets of the regularized objects and then take the 
limit $r\to 0$. The end result is the Lie algebra (\ref{2.10}) in the 
sense that $A(e)$ is identified with an element of Cyl$^\infty$ and 
$E_n(S)$ with the element $Y_n(S)\in$Vec. Thus, in terms of 
Poisson brackets we have $\{f,f'\}=0,\; \{E_n(S),f\}=Y_n(S)\cdot f,\; 
\{\{E_n(S),E_{n'}(S')\},f\}= [Y_n(S),Y_{n'}(S')]\cdot f$ for any 
$S,S',n,n'$ and $f,f'\in$Cyl$^\infty$. Notice that 
this implies that the subalgebra of fluxes becomes non-Abelian in 
apparent contradiction to the Abelian nature of the $D-$smeared objects 
$E(S_r)$. However, as one can show, we have still 
$\{\{E_n(S_r),E_{n'}(S'_r)\},f\}=0$, so the non-Abelianess comes about due 
to the singular smearing dimension. The reason for not using $D-$smeared
electric fields is that $\{E_n(S_r),f\}$ is no longer an element of 
Cyl (in the non-Abelian case), it is an integral over elements of 
Cyl and not a countable linear combination.\\
\\
The representation theory based on the abstract algebra $\cal P$ 
supplemented by appropriate $^\ast-$relations gets complicated due to 
the fact that the vector fields will be represented by unbounded 
operators, so that domain questions will arise. We avoid this by means of 
passing to the corresponding non-Abelian analogs of the Weyl elements.
\begin{Definition} \label{def2.4} ~~~~\\
For $t\in \mathbb{R}$ define
\be \label{2.11}
W^n_t(S):=e^{t Y_n(S)}=e^{-it [iY_n(S)]}
\ee
where $n^j(x) n^k(x)\delta_{jk}=1$. The algebra $\mathfrak{A}$ is defined 
as the 
free tensor algebra generated by the $(f,W^n_t(S))\in\mbox{Cyl}^\infty
\times \exp(Vec)$ modulo the two-sided ideal induced by (\ref{2.10}) and 
modulo the $^\ast-$relations 
\be \label{2.12a}
f^\ast=\overline{f},\;\;(W^n_t(S))^\ast=(W^n_{-t}(S))=(W^n_t(S))^{-1}
\ee
\end{Definition}
Due to the non-Abelian nature of the Group $G$, the relations in 
$\mathfrak{A}$ induced by 
(\ref{2.10}) are somewhat difficult to describe but it is 
nevertheless explicitly possible. In order to do this, we introduce the 
following notions. 
\begin{Definition} \label{def2.5} ~~~\\
i)\\
Let $x\in\Man$ be given. The germ $[e]_x$ of an edge $e$ with 
$b(e)=e(0)=x$
is defined by the infinite number of Taylor coefficients $e^{(n)}(0)$
in some parametrization. Likewise, the germ $[S]_x$ of a surface $S$ 
with $S(0,..,0)=x$ is defined by the Taylor coefficients 
$S^{(n_1,.,n_{D-1})}(0,..,0)$ in some reparametrization.\\
ii)\\ 
The set of germs $[e]$ ($[S]$) of edges (surfaces) at given $x\in\Man$ 
does not 
depend on $x$ and will be denoted by $\cal E$ ($\cal S$).\\  
iii)\\
Notice that the germs know about the orientation of $e,S$ and that 
their knowledge allows us to reconstruct $e(t),S(u_1,..u_{D-1})$ up to 
reparametrization due to analyticity so that they reconstruct $e,S$. We 
say that two germs are equal if they reconstruct the same edges and 
surfaces respectively from a given point $x$.\\
iv)\\
Let $x\in\Man,\;[e]\in {\cal E}$. We define 
elements $R^j_{x,[e]}\in$Vec by (assume w.l.g. that $\gamma$ is adapted 
to $x$ in the sense that each edge is either disconnected or or outgoing
from $x$)
\be \label{2.13}
R^j_{x,[e]} p_\gamma^\ast f_\gamma:=
p_{\gamma}^\ast \sum_{e'\in E(\gamma')} 
\delta_{x,b(e')} \delta_{[e],[e']} R^j_{e'} f_\gamma 
\ee
v)\\
Let $x\in\Man,\;[e]\in{\cal E},\;[S]\in {\cal S}$.
We define 
\be \label{2.14}
\sigma([S],[e]):=\sigma(S',e') \mbox{ for any } e',\;S'
\mbox{ s.t. } [e]=[e']_x,\;[S]=[S']_x
\ee
\end{Definition}
\begin{Lemma} \label{la2.1} ~~~~\\
i)\\
The vector fields $R^j_{x,[e]}$ satisfy the following commutation 
relations
\be \label{2.15}
{[}R^j_{x,[e]},R^k_{x',[e']}]=-f^{jk}\;_l \delta_{[e],[e']} 
\delta_{x,x'} R^l_{[x],[e]}
\ee
where $[\tau_j,\tau_k]=f_{jk}\;^l \tau_l$ defines the structure 
constants\footnote{Since $G$ is a compact, connected Lie group, we have 
$G/D\cong A\times S$ where $D$ is a central discrete subgroup and $A,S$ 
are Abelian and semisimple Lie 
groups respectively. Indices are 
dragged w.r.t. the Cartan-Killing metric Tr$(T_j T_k)=-\delta_{jk}$ where 
$(T_j)^k_l=f_{lj}\;^k,\;\;f_{jkl}$ totally skew for the semisimple
generators.}.\\
ii)\\
The flux vector fields $Y_n(S)$ can be expressed in terms of the 
$R^j_{x,[e]}$ by the formula
\be \label{2.16}
Y_n(S)=\sum_{x\in S} \sum_{[e]\in {\cal E}} \sigma([S]_x,[e]) 
n_j(x) R^j_{x,[e]}
\ee
\end{Lemma}
The proof of lemma \ref{la2.1} consists of a straightforward computation
in applying the left and right hand sides of (\ref{2.15}), (\ref{2.16})
to elements of Cyl$^\infty$. Formula (\ref{2.16}) looks cumbersome due to
the uncountably infinite sums involved, however, as vector fields on 
Cyl$^\infty$ they make perfect sense. Notice that (\ref{2.16}) allows us 
for the first time to compute the commutator $[Y^j_{S},Y^k_{S'}]$ in 
closed form and one sees immediately that the $Y_n(S)$ doe not form a 
subalgebra in Vec. The advantage of the germ vector fields is that they
have an extremely simple and closed algebra among themselves. They are not 
obviously generated from the $Y_n(S)$ and are thus not of physical 
interest, however, they are a useful tool in order to perform
practical calculations. We stress that only the algebra of vector fields 
generated
from the $Y_n(S)$ are of physical interest, but that algebra is 
a subalgebra of the bigger algebra generated by the $R^j_{x,[e]}$ and 
we may exploit that.

We can now compute the commutation relations among the $W^n_t(S)=\exp(t
Y_n(S))$ and the $f\in$Cyl$^\infty$. We have 
\ba \label{2.17} 
W^n_t(S) f (W^n_t(S))^{-1}&=&
\sum_{m=0^\infty} \frac{t^m}{m!} [Y_n(S),f]_{(m)}
\nonumber\\
&=&
\sum_{m=0^\infty} \frac{t^m}{m!} (Y_n(S))^m f
=W^n_t(S)\cdot f
\ea
where the bracket notation denotes the multiple commutator and the last 
line denotes the application of the exponentiated vector field to a 
cylindrical function. Let now $f=p_\gamma^\ast f_\gamma$. Since the 
$R^j_e$ are mutually commuting we have 
\ba \label{2.18}
[W^n_t(S)\cdot f ](A)
&=& [p_\gamma^\ast \prod_{e\in E(\gamma)} e^{t \sigma(S,e) n_j(b(e)) 
R^j_e} 
f_\gamma](A)
\nonumber\\
&=& f_\gamma(\{e^{t \sigma(S,e) n^j(b(e)) \tau_j} A(e)\}_{e\in E(\gamma)})
\ea
To see the equality in the last line of (\ref{2.18}) it is 
obviously sufficient to show it for one copy of $G$, that is
\be \label{2.19}
f_t(h):=[e^{t n_j R^j} f](h)=f(e^{t n^j \tau_j}h)=:f'_t(h)
\ee
for $f\in C^\infty(G)$ and any $t\in \Rl$. To show this, set 
$R:=n_j R^j,\; \tau:=n^j\tau_j$. We clearly have $f_0(h)=f'_0(h)=f(h)$
and 
\ba \label{2.20}
(\frac{d}{dt}f_t)(h) &=& [\frac{d}{ds}]_{s=0} (e^{(s+t) R}f)(h)
=[\frac{d}{ds}]_{s=0} (e^{sR}\; e^{tR}f)(h)
=(R\; e^{tR}f)(h)=(R f_t)(h)
\nonumber\\
(\frac{d}{dt}f'_t)(h) &=& [\frac{d}{ds}]_{s=0} f(e^{(s+t) \tau}h)
=[\frac{d}{ds}]_{s=0} f(e^{s\tau}\; e^{t\tau}h)
=(R\; f)(e^{t\tau} h)=(R f'_t)(h)
\ea
Hence, $f_t,f'_t$ satisfy the same ordinary differential equation and 
initial conditions and thus (\ref{2.19}) follows from the uniqueness and 
existence theorems about ordinary differential equations. 
Hence, the $W^n_t(S)$ act on cylindrical functions just by left 
translation in their arguments as was to be expected and 
the result (\ref{2.20}) implies that the algebra $\mathfrak{A}$ can be 
extended to the bounded cylindrical functions Cyl$_b$ on $\a$ 
(differentiability is no longer necessary) which forms an Abelian 
subalgebra.

Finally we compute the commutator of Weyl-operators by explicitly using 
the germ vector fields. We have 
\be \label{2.21}
W^n_t(S) W^{n'}_{t'}(S') (W^n_t(S))^{-1}
=\exp(t' n'_j \sum_{x'\in S'}\sum_{[e']\in{\cal E}} \sigma([S']_{x'},[e'])
W^n_t(S) R^j_{x',[e']} W^n_t(S)^{-1})
\ee
Now 
\be \label{2.22}
W^n_t(S) R^j_{x',[e']} (W^n_t(S))^{-1}
= \sum_{m=0}^\infty \frac{t^m}{m!} 
[\sum_{x\in S}\sum_{[e]\in {\cal E}} \sigma([S]_x,[e]) n_k(x) R^k_{x,[e]},
R^j_{x',[e']}]_{(m)}
\ee
and 
\ba \label{2.23}
&& [\sum_{x\in S}\sum_{[e]\in {\cal E}} \sigma([S]_x,[e]) n_k(x) 
R^k_{x,[e]},
R^{j_1}_{x',[e']}]_{(m)}
\nonumber\\
&=&[-f_{k_1 j_1 j_2} \sigma([S]_{x'},[e']) \chi_S(x') n^{k_1}(x')]
[\sum_{x\in S}\sum_{[e]\in {\cal E}} \sigma([S]_x,[e]) n_k(x) R^k_{x,[e]},
R^{j_2}_{x',[e']}]_{(m-1)}
\nonumber\\
&=& \sigma([S]_{x'},[e']) \chi_S(x') n_{j_1 j_2}(x')
[\sum_{x\in S}\sum_{[e]\in {\cal E}} \sigma([S]_x,[e]) n_k(x) R^k_{x,[e]},
R^{j_2}_{x',[e']}]_{(m-1)}
\nonumber\\
&=& \sigma([S]_{x'},[e'])^2 \chi_S(x') (n^2)_{j_1 j_2}(x')
[\sum_{x\in S}\sum_{[e]\in {\cal E}} \sigma([S]_x,[e]) n_k(x) R^k_{x,[e]},
R^{j_2}_{x',[e']}]_{(m-2)}
\nonumber\\
&=& \sigma([S]_{x'},[e'])^m \chi_S(x') (n^m)_{j_1 j_2}(x')
R^{j_2}_{x',[e']}
\ea
where we have defined the matrix $n$ by 
$n_{jj'}:=n^k f_{jkj'}$ and 
$\chi_S$ is the characteristic function of the set $S$. Hence (\ref{2.22})
becomes
\ba \label{2.24}
&& W^n_t(S) n'_j(x') R^j_{x',[e']} (W^n_t(S))^{-1}\nonumber\\
&=&
{[}1-\chi_S(x')]n'_j(x') R^j_{x',[e']}
+\chi_S(x') n'_j(x') \sum_{m=0}^\infty 
\frac{[\sigma([S]_{x'},[e'])t]^m}{m!}
(n^m)_{j k}(x') R^k_{x',[e']}
\nonumber\\
&=& [1-\chi_S(x')]n'_j(x') R^j_{x',[e']}
+\chi_S(x') 
[e^{-t \sigma([S]_{x'},[e']) \mbox{ad}_n(x')}]_{jk} 
R^k_{x',[e']}
\ea
where for any $\tau\in$Lie$(G)$ the matrix ad$_\tau$ is defined by 
$[\mbox{ad}_\tau]_{jk} \tau_k=[\tau,\tau_j]$. Plugging (\ref{2.24}) into
(\ref{2.21}) we obtain
\ba \label{2.25}
&&W^n_t(S) W^{n'}_{t'}(S') (W^n_t(S))^{-1}
\\
&=& \exp(t' \sum_{[e]\in{\cal E}} 
[\sum_{x\in S'-S} n'_j(x)\sigma([S']_x,[e]) \delta_{jk} 
+\sum_{x\in S\cap S'} n'_j(x) \sigma([S']_x,[e])
[e^{-t \sigma([S]_{x},[e]) \mbox{ad}_n(x)}]_{jk} 
R^k_{x,[e]}]) \nonumber
\ea
which is almost of the form of a $W^t_n(S)$ again. That the $W^n_t(S)$ do 
not close among each other we knew from the associated statement for the 
$Y^j_S$, however, we see that the algebra they generate can be computed 
explicitly.\\
\\
Finally we equip the algebra $\mathfrak{A}$ with a $C^\ast$ structure.
Since the operator norm in a representation $\pi$
of $\mathfrak{A}$ on a Hilbert space $\cal H$ {\it does} define a $C^\ast-$
norm through $||a||:=||\pi(a)||_{{\cal H}}$ we just need to find a 
representation of $\mathfrak{A}$ to close our algebra to a $C^\ast$
algebra. However, the Ashtekar -- Lewandowski 
Hilbert space ${\cal H}_0$ {\it is} a representation space for a 
representation $\pi_0$, hence such a $C^\ast$-norm exists.\footnote{We
  note however the possibility that one can find a different
  $C^\ast$ closure of the algebra at hand. Since we will not use any
  specific information about the closure, however, this is of no
  concern in the present paper.}

That representation is given
by ${\cal H}_0=L_2(\ab,d\mu_0)$ where $\ab$ is the spectrum of the 
$C^\ast-$subalgebra of $\mathfrak{A}$ given by Cyl and $\mu_0$ is a 
regular Borel probability measure on $\ab$ consistently defined by
\be \label{2.26}
\mu_0(p_\gamma^\ast f_\gamma)=\int_{G^{|E(\gamma)|}} 
\prod_{e\in E(\gamma)} d\mu_H(h_e) f_\gamma(\{h_e\}_{e\in E(\gamma)})
\ee
for measurable $f_\gamma$ and extended by $\sigma-$additivity. 
Then
\be \label{2.27}
\pi_0(f) \psi=f(A)\psi \mbox{ and } 
\pi_0(W^n_t(S)) \psi=W^n_t(S) \psi 
\ee
so that Cyl$_b$ is represented by bounded multiplication operators 
while the Weyl elements $W^n_t(S)$ are simply extended from Cyl$_b$ to 
$L_2(\ab,d\mu_0)$. Notice that Cyl$_b$ (in particular the 
continuous functions Cyl$^0$ on $\ab$) are dense in ${\cal H}_0$ because 
${\cal H}_0$ is the GNS Hilbert space induced by the positive linear 
functional $\omega_0$ on Cyl$_b$ defined by $\omega_0(f)=\mu_0(f)$.
Finally it follows from the left invariance of the Haar measure that 
$\pi_0(W^n_t(S)$ are unitary operators as they should be. Thus 
e.g. 
\ba \label{2.28}
||f||_{\mathfrak{A}}&=&||\pi_0(f)||_{{\cal B}({\cal H}_0)}
=\sup_{||\psi||=1} ||f\psi||_{{\cal H}_0}=\sup_{a\in \a} |f(A)|
\nonumber\\
||W^n_t(S)||_{\mathfrak{A}}&=&||\pi_0(W^n_t(S))||_{{\cal B}({\cal H}_0)}
=\sup_{||\psi||=1} ||W^n_t(S)\psi||_{{\cal H}_0}=1
\ea
where $\cal B$ denotes the bounded operators on a Hilbert space. The 
$C^\ast-$norm of any other element of $\mathfrak{A}$ can be computed by 
using the commutation relations and the inner product on ${\cal H}_0$.\\
\\
This concludes our exposition about the $C^\ast-$algebra $\mathfrak{A}$.

\section{General Representation Theory of $\mathfrak{A}$}
\label{s3}
Let us clarify what we mean by a representation of $\mathfrak{A}$. 
\begin{Definition} \label{def3.1} ~~~~
By a representation of $\mathfrak{A}$ we mean an $^\ast-$algebra 
homomorphism 
$\pi:\;\mathfrak{A}\to {\cal B}({\cal H})$ from $\mathfrak{A}$ into
the algebra of bounded operators of a Hilbert space $\cal H$. Thus
$\pi(a+zb)=\pi(a)+z\pi(b),\;\pi(ab)=\pi(a)\pi(b),\;\pi(a^\ast)=
[\pi(a)]^\dagger$ for all $a,b\in\mathfrak{A},z\in\Cl$.
\end{Definition}
The representation theory of $\mathfrak{A}$ is very rich and first steps 
towards a classification have been made in \cite{10}. An elementary result
is the following.
\begin{Lemma} \label{la3.1} ~~~\\
The representation space $\cal H$ of $\mathfrak{A}$ is necessarily a 
direct sum of Hilbert spaces
\be \label{3.1}
{\cal H}=\oplus_\nu {\cal H}_\nu
\ee
where ${\cal H}_\nu=L_2(\ab,d\mu_\nu)$ is an $L_2$ space over 
the spectrum $\ab$ of Cyl$_b$, and $\mu_\nu$ is a probability
measure on $\ab$.
\end{Lemma}
Proof of lemma \ref{la3.1}:\\
Every representation $\pi$ of $\mathfrak{A}$ on a Hilbert space ${\cal 
H}$ is, in particular, a representation of the Abelian sub$^\ast-$algebra 
Cyl$_b$. Now a general result from $C^\ast-$algebra theory \cite{20}
says that every non-degenerate representation (that is, 
Ker$(\pi):=\{\psi\in{\cal H};\;\pi(a)\psi=0\;\forall\;a\in\mathbb(A)\}=\{0\}$)
is a direct sum of cyclic
representations $\pi_\nu$ on Hilbert spaces ${\cal H}_\nu$. Now 
since our algebra is unital $1\in \mbox{Cyl}_b\subset \mathfrak{A}$ we 
necessarily have $\pi(1)=\mbox{id}_{{\cal H}}$, hence $\pi$ is 
non-degenerate and also its restriction to Cyl$_b$ is. Let $\Omega_\nu$
be a unit vector in ${\cal H}_\nu$ which is cyclic for Cyl$_b$. Define 
the normalized, positive linear functional on Cyl$_b$ given by
\be \label{3.2}
\omega_\nu(f):=<\Omega_\nu,\pi_\nu(f)\Omega_\nu>_{{\cal H}_\nu}
\ee
Since Cyl$_b$ is an Abelian $C^\ast-$algebra, by Gel'fands theorem,
we may think of it as the algebra of continuous functions $C(\ab)$
on the Gel'fand spectrum $\ab$ of Cyl$_b$. Since $\ab$ is a compact 
Hausdorff space, by the Riesz representation theorem, the positive 
linear functional $\omega_\nu$ uniquely determines a regular Borel 
probability measure $\mu_\nu$ on $\ab$ via 
\be \label{3.3}
\omega_\nu(f)=\int_{\ab} d\mu_\nu(A) f(A)
\ee
hence we may choose w.l.g. $\Omega_\nu=1,\; 
\pi_\nu(f)=f\mbox{id}_{{\cal H}_\nu}$ (multiplication
operator) and ${\cal H}_\nu=L_2(\ab,d\mu_\nu)$ (by 
the GNS construction, any
other choice corresponds to a unitary transformation).\\
$\Box$\\
A generic element $\psi\in {\cal H}$ is therefore given by 
\be \label{3.4}
\psi=\oplus_\nu \psi_\nu=\sum_\nu \psi_\nu 1^\nu
\ee
where $\psi_\nu\in{\cal H}_\nu$, $1^\nu_{\nu'}=\delta^\nu_{\nu'}\cdot 1$
and $\sum_\nu ||\psi_\nu||_\nu^2<\infty$. Here we have denoted the 
norm on ${\cal H}_\nu$ by $||.||_\nu$.

It follows that 
\be \label{3.5}
\pi(f)\psi=\oplus_\nu f\psi_\mu=f\oplus_\nu \psi_\nu=f \psi
\ee
whence 
\be \label{3.6}
\pi(f)=f\mbox{id}_{{\cal H}}
\ee
is simply a multiplication operator on ${\cal H}$. 

Notice that while the subalgebra Cyl$_b$ 
has no off-diagonal entries, that is in general not the case for the 
$\pi(W^n_t(S)$. Also, while $\pi_\nu$ is a cyclic representation 
for Cyl$_b$, $\pi$ is not necessarily cyclic for Cyl$_b$, one will 
generically assume it to be cyclic for the full algebra $\mathfrak{A}$
only (that is, there is a vector $\Omega\in {\cal H}$ such that 
the set of states given by $\pi(a)\Omega,\;a\in \mathfrak{A}$ is dense in 
$\cal H$). In what 
follows we will only consider representations which 
are cyclic for $\mathfrak{A}$ (otherwise we can decompose $\pi$ further 
into cyclic representations by the above theorem, hence cyclic 
representations are the basic building blocks). 

This all that one can 
say so far about general representations of $\mathfrak{A}$ without making
further assumptions. To get further structural control over the 
representation theory one must examine restricted situations of physical
interest. In the next section we will
study the important class of diffeomorphism invariant representations
which are those realized in nature (nature {\it is} diffeomorphism 
invariant, so there is no need to study other representations at 
all, at least from a physics point of view).

\section{Diffeomorphism Invariant Representations of $\mathfrak{A}$ 
and a Uniqueness Theorem}
\label{s4}

The group Diff$^\omega(\Man)$ of analytic 
diffeomorphisms on $\Man$ 
has a natural representation as 
outer automorphisms
on $\mathfrak{A}$ defined for any $\varphi\in$Diff$^\omega(\Man)$ 
by
\ba \label{4.1}
\alpha_\varphi(p_\gamma^\ast f_\gamma)&=& 
p_{\varphi^{-1}(\gamma)}^\ast f_\gamma
\nonumber\\
\alpha_\varphi(W^n_t(S))&=& W^{n\circ \varphi}_t(\varphi^{-1}(S))
\ea 
and extended by the automorphism property 
$\alpha_\varphi(ab)=\alpha_\varphi(a)\alpha_\varphi(b),\;
\alpha_\varphi(a+zb)=\alpha_\varphi(a)+z\alpha_\varphi(b)$. It is trivial
to check that 
$\alpha_\varphi\circ \alpha_{\varphi'}=\alpha_{\varphi\circ\varphi'}$.

Likewise, the set Fun$(\Man,G)$, which forms a group under pointwise 
multiplication, has a natural representation as 
outer automorphisms
on $\mathfrak{A}$ defined for any $g\in$Fun$(\Man,G)$ 
by
\ba \label{4.1a}
[\alpha_g(p_\gamma^\ast f_\gamma)](A)&=& 
f_\gamma(\{g(b(e))A(e)g(f(e))^{-1})
\nonumber\\
\alpha_g(W^n_t(S))&=& W^{n^g}_t(S)
\ea 
where $n^g(x)=\mbox{Ad}_{g(x)}(n(x)),\;n(x)=n_j\tau_j$. As one can check, 
with these definitions we have $\alpha_\varphi\circ \alpha_g\circ 
\alpha_{\varphi^{-1}}=\alpha_{\varphi^\ast g}$ so that the combined 
kinematical gauge group acquires the structure of a semidirect 
product ${\cal G}=\mbox{Fun}(\Man,G)\rhd \mbox{Diff}^\omega(\Man)$
if we define $\alpha_{(g,\varphi)}:=\alpha_g\circ\alpha_\varphi$ with
Fun$(\Man,G)$ as invariant subgroup.
\begin{Definition} \label{def4.1} ~~~\\
i)\\
A cyclic representation $\pi$ of $\mathfrak{A}$ is said to be diffeomorphism
invariant provided that there is a unitary 
representation 
\be \label{4.2}
U_\pi:\;\mbox{Diff}^\omega(\Man)\to {\cal B}({\cal H});\;
\varphi\mapsto U_\pi(\varphi)
\ee
of the diffeomorphism group and a cyclic invariant vector 
$\Omega\in {\cal H}$ such that 
\be \label{4.3}
U_\pi(\varphi) \pi(a) U_\pi(\varphi)^{-1}=\pi(\alpha_\varphi(a))
\mbox{ and } U_\pi(\varphi)\Omega=\Omega
\ee
for all $a\in \mathfrak{A},\; \varphi\in$Diff$^\omega(\Man)$.
There are similar definitions of gauge invariant or kinematically 
invariant representations when replacing Diff$^\omega(\Man)$ by
Fun$(\Man,G)$ or $\cal G$.\\
ii)\\
Consider each $\psi\in{\cal H}$ as a vector-valued function of
$A\in \ab$ according to 
\be \label{4.3a}
\psi(A):=\oplus_\nu \psi_\nu(A) 
\ee
The {\bf natural} (pull-back) representation of Diff$^\omega(\Man)$
is defined by 
\be \label{4.3b}
U_\pi(\varphi)\psi := \oplus_\nu \alpha_\varphi(\psi_\nu)
\ee
Likewise the {\bf natural} representation of Fun$(\Man,G)$ is defined by
\be \label{4.3c}
U_\pi(g)\psi := \oplus_\nu \alpha_g(\psi_\nu)
\ee
\end{Definition}
The name {\it natural} representation is due to the fact that it is the 
natural lift of the action of 
diffeomorphisms or gauge transformations on functions of $\a$, that is 
$f(\varphi^\ast A)\cong[\alpha_\varphi(f)](A)$ and 
$f(A^gA)\cong[\alpha_g(f)](A)$
to functions of $\ab$.
The natural representation has the feature of leaving constant functions 
invariant, which are therefore natural candidates for cyclic invariant 
vectors.

A natural starting point for cyclic invariant representations exists, 
provided one manages to find a positive linear functional $\omega$
on $\mathfrak{A}$ with the invariance property
\be \label{4.4} 
\omega(\alpha_\varphi(a))=\omega(a)
\ee
for all $a\in \mathfrak{A},\; \varphi\in$Diff$^\omega(\Man)$. Namely, 
let $\pi_\omega,\Omega_\omega,{\cal H}_\omega$ be the GNS data for 
$\omega$ \cite{20}, that is, 
\be \label{4.5}
\Omega_\omega:=[1],\;
\pi_\omega(a)\Omega_\omega:=[a],\;
<\pi_\omega(a)\Omega_\omega,\pi_\omega(b)\Omega_\omega>:=\omega(b^\ast a)
\ee
where $[a]$ is the equivalence class $\{a+b;\;\omega(b^\ast b)=0\}$.
Then 
\be \label{4.6}
U_\omega(\varphi)\pi_\omega(a)\Omega_\omega:=
\pi_\omega(\alpha_\varphi(a))\Omega_\omega
\ee
represents the diffeomorphism group unitarily as inner automorphisms of
${\cal B}({\cal H}_\omega)$ with $\Omega_\omega$ as cyclic invariant 
vector.

Using the language of the present paper, in \cite{10} the following result 
was established.
\begin{Theorem} \label{th4.1} ~~~~\\
Suppose that \\
1) $G=U(1)$.\\
2) $\pi$ is cyclic already for Cyl$_b$ so that necessarily 
${\cal H}=L_2(\ab,d\mu)$ with cyclic vector $\Omega=1$ by lemma 
\ref{la3.1}.\\
3) $\pi$ is diffeomorphism invariant with $\Omega$ as invariant cyclic 
vector where the diffeomorphisms act by pull back.\\
4) The one parameter subgroups $t\mapsto \pi(W^n_t(S))$ are weakly 
continuous.\\
5) $\Omega$ is in the domain of any self-adjoint generator 
$-i[\frac{d}{dt}]_{t=0} \pi(W^n_t(s))$.\\
Then necessarily ${\cal H}=L_2(\ab,d\mu_0)={\cal H}_0$ is the 
Ashtekar -- Lewandowski representation.
\end{Theorem}
Several of the assumptions of theorem \ref{th4.1} are unsatisfactory:
First of all, the restriction to $U(1)$ makes it of limited physical 
relevance since in particular loop quantum gravity would need such a 
result for general compact groups. Next, it is not natural to require 
that already Cyl$_b$ is cyclic for the representation, the most general 
interesting representations will be those for which only the full algebra 
$\mathfrak{A}$ is cyclic. Furthermore, while it is natural to assume that 
the constants are in the domain of the self-adjoint generators of the 
Weyl elements (because the unit function is a cyclic vector for 
Cyl$_b$), one has 
no intuition whether there are not more general
representations which violate this assumption. 

On the other hand, if one does not assume weak continuity of the fluxes 
then the requirements will be too weak to limit the number of possible 
representations. This is already the case for the Schr\"odinger 
representation of ordinary quantum mechanics: If one gives up weak
continuity of the Weyl elements then many more representations exist
which are not captured by the Stone -- von Neumann theorem. In fact,
the Stone -- von Neumann theorem not only requires the representation to 
be cyclic but even to be irreducible (that is, every vector is cyclic), 
otherwise also more representations result. We thus expect to find a 
strong result also only in the irreducible case. Irreducibility is 
actually more physical than cyclicity since then no non-trivial invariant
subspaces exist and moreover, there are no distinguished cyclic elements.
We do not know at present whether  
cyclicity is actually enough for the result to be proved below.

There is one more unnatural assumption in theorem \ref{th4.1}: Why should 
it be the case that the vector $1$ is left invariant by $U_\pi(\varphi)$ ?
If we have only cyclicity of $\mathfrak{A}$ then it is also not clear 
why it should be the vector $1$ which is cyclic. Actually, this 
discussion leads to the representation theory of Diff$^\omega(\Man)$ as
the following discussion reveals:\\
Suppose that we do have a diffeomorphism invariant representation
$\pi$. Then the action of $U_\pi(\varphi)$ is known on the whole 
representation space ${\cal H}$ provided we know it on the $1^\nu$ because 
\be \label{4.7}
U_\pi(\varphi)\psi=U_\pi(\varphi)\sum_\nu \pi(\psi_\nu) 1^\nu
=\sum_\nu \pi(\alpha_\varphi(\psi_\nu)) U_\pi(\varphi) 1^\nu
\ee
The natural pull -- back representation of definition \ref{def4.1} would 
assign $U_\pi(\varphi) 1^\nu=1^\nu$. But can we say more about the 
possible representations $U_\omega(\varphi)$ ?\\
Suppose that we are interested in asymptotically flat situations. Then,
if we include among Diff$^\omega(\Man)$ also symmetries of the 
asymptotically Minkowskian metric, then Diff$^\omega(\Man)$ will contain
the asymptotic Poincar\'e group as a subgroup and one concludes that 
$U_\pi$ is in particular a unitary representation of the 
asymptotic Poincar\'e group. The {\it continuous}\footnote{Here we mean 
continuity of one parameter subgroups.}, 
irreducible, unitary 
representations of that group have been classified by Wigner and it seems 
that one is in a good position. However, it is unclear if and how these 
representations can be extended to all of Diff$^\omega(\Man)$ and,
moreover, it turns out that generic representations will even violate 
the continuity assumption: For instance, the pull-back representation 
of one -- parameter subgroups of Diff$^\omega(\Man)$ on the Ashtekar -- 
Lewandowski space ${\cal H}_0$ is not weakly continuous \cite{1}.

To see that there really is an abundance of unitarily inequivalent 
representations of Diff$^\omega(\Man)$, suppose that we start from a
a representation $\pi$ of $\mathfrak{A}$ on a Hilbert space $\cal H$ with 
unitary pull-back representation $U_\pi$ of Diff$^\omega(\Man)$. Let 
$W\in {\cal B}({\cal H})$ be any bounded operator with bounded inverse 
which we consider as being of the form $W=\pi(a)$ for some 
$a\in \mathfrak{A}$. Let us also denote 
$\alpha_\varphi(W):=\pi(\alpha_\varphi(a))$. We claim that
\be \label{4.8}
U'_\pi(\varphi):=W^{-1}\alpha_\varphi(W)U_\pi(\varphi)
\ee
defines a representation of Diff$^\omega(\Man)$ on $\cal H$. We have
\ba \label{4.9}
U'_\pi(\varphi)U'_\pi(\varphi')  
&=&W^{-1}\alpha_\varphi(W)U_\pi(\varphi)
W^{-1}\alpha_{\varphi'}(W)U_\pi(\varphi')
\nonumber\\
&=& W^{-1}\alpha_\varphi(W)[U_\pi(\varphi)W^{-1} U_\pi(\varphi)^{-1}]
[U_\pi(\varphi) \alpha_{\varphi'}(W) U_\pi(\varphi)^{-1}]
U_\pi(\varphi) U_\pi(\varphi')
\nonumber\\
&=& W^{-1}
\alpha_\varphi(W)\alpha_\varphi(W^{-1})
\alpha_\varphi(\alpha_{\varphi'}(W))
U_\pi(\varphi\circ \varphi')
\nonumber\\
&=& W^{-1}\alpha_{\varphi\circ\varphi'}(W)
U_\pi(\varphi\circ \varphi')=U'_\pi(\varphi\circ\varphi')
\ea
Since the two representations are equivalent,
\be \label{4.10}
WU'_\pi(\varphi) W^{-1}=U_\pi(\varphi)
\ee
the requirement that also $U'_\pi(\varphi)$ is a unitary representation
leads to the condition
\be \label{4.11}
\alpha_\varphi(WW^\dagger)=WW^\dagger
\ee
and can be satisfied, for instance, if $WW^\dagger$ is a constant matrix.
In that case the polar decomposition gives $W=CV$ where $C$ is an arbitrary 
constant self-adjoint and positive operator while $V$ is unitary.  
Of course, the question is whether the representation $U'_\pi$ on $\cal H$ 
is unitarily equivalent to $U_\pi$. As (\ref{4.11}) reveals,
this will be the case if and only if $W$ is a unitary operator, that is,
$C=\mbox{id}_{{\cal H}}$.

One might think that one can bring more structure into the analysis 
by requiring that the representation 
$U_\pi$ to be irreducible as well (not only $\pi$) because then it follows 
from
Schur's lemma that $W=\lambda\mbox{id}_{{\cal H}}$ and unitary equivalence
requires $|\lambda|=1$. However, it is well known that 
interesting representations 
of the diffeomorphism group are generically quite reducible. For instance,
the pull back representation on ${\cal H}_0$ is extremely reducible 
\cite{1}, we have a countably (under suitable superselection 
criteria \cite{21}) infinite direct sum decomposition
\be \label{4.12}
{\cal H}_0=\oplus_{[\gamma]} {\cal H}^{[\gamma]}_0
\ee
where the non-separable, mutually orthogonal, invariant subspaces 
${\cal H}^{[\gamma]}_0$ are spanned by spin network functions 
\cite{22} over graphs 
$\gamma'$ which belong to one and same (generalized) knot class 
$[\gamma]$. None of these subspaces alone captures interesting physics 
and thus it seems to be an unphysical requirement to restrict to 
irreducible representations of Diff$^\omega(\Man)$.

These cautionary remarks are just to indicate that there are a priory many 
inequivalent, unitary representations of Diff$^\omega(\Man)$ available
and their classification goes beyond the scope of the present paper.
The selection of one of them might be comparable to the selection of a definite
spin representation of the Poincar\'e group, however, {\it it is much 
more complicated} (the diffeomorphism group is an infinite dimensional
group !). Accordingly, we must be modest and specify the representation 
of Diff$^\omega(\Man)$ in the statement of our theorem below. 
Obviously, we will choose the pull-back representation which is natural 
because it is available in any representation of $\mathfrak{A}$ as shown
in lemma \ref{la3.1}. 

Before we state our theorem, let us define the notion of a spin network 
function on $\ab$.
\begin{Definition} \label{def4.2} ~~~\\
Choose precisely on representative $\rho$ from each 
equivalence class of irreducible representations of $G$, denote 
by $d_\rho$ the dimension of the representation space of $\rho$ and 
denote for any $h\in G$ and $M,N=1,..,d_\rho$ by $\rho_{MN}(h)$ the matrix
elements of the unitary matrix $\rho(h)$. Consider a graph $\gamma$ 
together with a labeling of each of its edges $e\in E(\gamma)$ 
with label $\rho_e,M_e,N_e;\;M_e,N_e=1,..,d_{\rho_e}$ and collect
them into a {\bf spin network} 
\be \label{4.12a}
s=(\gamma,\;
\vec{\rho}=\{\rho_e\}_{e\in E(\gamma)},\;
\vec{M}=\{M_e\}_{e\in E(\gamma)},\;
\vec{N}=\{N_e\}_{e\in E(\gamma)})
\ee
Then the {\bf spin network function} with label $s$ is given by
\be \label{4.12b}
T_s(A)=\prod_{e\in E(\gamma)}\;
\{\sqrt{d_{\rho_e}}\;[\rho_e(A(e))]_{M_e N_e}\}
\ee
\end{Definition}
As one can show, they provide an orthonormal basis for ${\cal H}_0$.\\
\\
We can then state our main result.
\begin{Theorem} \label{th4.2} ~~~\\
Let $G$ be a compact, connected gauge group, $\Man$ an oriented 
$D-$manifold 
with fixed analytic structure and associated diffeomorphism group 
Diff$^\omega(\Man)$. Let $\mathfrak{A}$ be the Weyl algebra generated 
from bounded cylindrical functions on the space $\a$ of smooth 
$G-$connections over $\Man$ and and exponentiated electric fluxes.\\
Let $\pi$ be a representation of $\mathfrak{A}$ on a Hilbert space 
${\cal H}$ with corresponding representation $U_\pi$ of 
Diff$^\omega(\Man)$.\\
Suppose that\\
i) {\bf Irreducibility}\\
$\pi$ is irreducible.\\
ii) {\bf Continuity}\\
The one -- parameter groups $t\mapsto \pi(W^n_t(S))$ are weakly 
continuous.\\
iii) {\bf Diffeomorphism Invariance}\\
The representation $U_\pi$ of Diff$^\omega(\Man)$ is 
unitary and coincides with the natural representation.\\
\\
If $G$ is not Abelian we also must require\footnote{At least presently.
Our results for the Abelian case indicate that this condition can be 
abolished, however, we were not able to circumvent it for now.}:\\
iv) {\bf Domains}\\
The vectors $1^\nu$ are in the common dense domain of the operators 
$\pi(Y_n(S)),\;\pi(Y_n(S))\pi(Y_{n'}(S))$ for any $S,n,n'$. 
\\
Then $\pi$ is unitarily equivalent to the Ashtekar -- Lewandowski
representation $\pi_0$.
\end{Theorem}
As we have discussed above, the only weak (that is, possibly overly 
restrictive) assumption in this theorem left is, as compared to 
theorem \ref{th4.1}, that we 
restrict ourselves to 
the natural representation of the diffeomorphism group (and the 
natural representation of the gauge group in the non-Abelian case). In 
particular,
there is no longer a restriction to a particular gauge group, to 
particular domains or to particular subalgebras which should already be 
cyclic.\\
\\
Proof of theorem \ref{th4.2}:\\
Before we go into the technical details, let us explain the strategy 
of the proof:\\
{\it Step 1: Continuity $\Rightarrow$ Dense Domain 
$\cal D$ of Individual Fluxes}\\
The self-adjoint generators $\pi(E_n(S))$ of the 
unitary groups $t\mapsto \pi(W^n_t(S))$ defined 
by 
\be \label{4.13}
-i\pi(E_n(S))=[\frac{d}{dt}]_{t=0} \pi(W^n_t(S))
\ee
which exist due to our continuity 
assumption, are not everywhere defined,
however, it turns out that one can always find, without additional 
assumptions, a suitable dense domain 
${\cal D}^n_S$ of $\pi(E_n(S))$ on which we will be able to work out the 
consequences of 
diffeomorphism invariance and unitarity of the Weyl elements.\\
{\it Step 2: Analytic Diffeomorphisms}\\
This step is technical and prepares for step 3 in which existence and 
properties of certain analytic diffeomorphisms are needed. If the 
assumptions i), ii), iii) and iv) displayed in theorem \ref{th4.2} 
are supplemented by an assumtion v) about the gauge group then this step
can be avoided as we show in appendix \ref{sa}.\\
{\it Step 3: Unitarity + Diffeomorphism Invariance $\Rightarrow$ 
$\mu_\nu=\mu_0$}\\
The domain ${\cal D}^n_S$ contains (multiples of) the so-called 
spin-network
functions \cite{22}. These are labeled, among other things, by a graph 
$\gamma$ and an irreducible representation $\pi_e$ for each edge 
$e\in E(\gamma)$. One can 
show that linear combinations of those are 
eigenfunctions of (polynomials of) the $\pi(E_n(S))$ with an eigenvalue 
which depends on the $\pi_e$ and on the number of intersections $N_e$
of $S$ with $e$. Using unitarity of the Weyl elements or the 
corresponding symmetry of the fluxes one can establish a relation
between this eigenvalue, which diverges as $N_e\to \infty$, and 
the expectation value of the spin network functions (considered
as bounded operators on ${\cal H}_\nu$), which is bounded, because due to 
diffeomorphism 
invariance that expectation value is actually independent of $N_e$.  
This leads to a contradiction unless $\mu_\nu=\mu_0$ is the 
Ashtekar -- Lewandowski measure. \\
{\it Step 4: Diffeomorphism Invariance + Irreducibility $\Rightarrow$
${\cal H}={\cal H}_0$}\\
The results of step 3 do not yet exclude the possibility that 
$\pi(W^n_t(S))$ has off-diagonal action on the direct sum of 
the ${\cal H}_\nu\cong {\cal H}_\nu$. Making use of our already available 
knowledge that all the Hilbert spaces are Ashtekar Lewandowski Hilbert 
spaces with explicitly known spin network basis, it is possible to show 
that $\pi(W^n_t(S))=W^n_t(S)\mbox{id}_{{\cal H}}$ is diagonal and 
all entries are equal. This contradicts the irreducibility condition 
unless there is only one copy of ${\cal H}_0$.\\
\\
From the structure of the proof it is clear that all assumptions are 
used in an essential way, in particular, there would be no result
if weak continuity is given up. Let us now go to the details.\\
\\
{\bf Step 1:}\\
\\
Let us write $S_n:=(n,S)$ in what follows and similarly 
$W_t(S_n):=W^n_t(S)$.
The following trick for how to construct a dense domain for all
the fluxes follows the proof of Stone's theorem that establishes a
one -- to one correspondence between self-adjoint operators on a Hilbert 
space and weakly continuous one parameter unitary groups.
\begin{Lemma} \label{la4.1}  ~~~~\\
Let $\phi\in C^\infty_c(\Rl)$
be a smooth test function 
For any $\psi\in {\cal H}$ and any $S_n$ define 
\be \label{4.14}
\psi_{\phi,S_n}:=\int_{\Rl} dt \phi(t) \pi(W_t(S_n))\psi
\ee
Then for each $S_n$ the finite linear combinations of the vectors 
$\pi(T_s) 1^\nu_{\phi,S_n}$ form a dense set ${\cal D}(S_n)$ in 
$\cal H$ as $s,\;\phi,\;\nu$ vary.
\end{Lemma}
Proof of lemma \ref{la4.1}:\\
The functions $C(\ab)$ are dense in ${\cal H}_\nu=L_2(\ab,d\mu_\nu)$
for any $\nu$. Finite linear combinations of spin network functions 
(which form a unital $^\ast-$subalgebra of Cyl$^\infty\subset C(\ab)$ which 
separates the points of $\ab$)
are dense in $C(\ab)$ with respect to the sup norm $||.||_\infty$ on 
$C(\ab)$ by 
the Weierstrass theorem.
Since $||.||_2\le ||.||_\infty$ it follows that finite linear combinations 
of spin network functions are dense in any ${\cal H}_\nu$, hence finite
linear combinations of the functions $\pi(T_s) 1^\nu$ are dense in 
${\cal H}$ as $s,\;\nu$ vary. 

Now for any $\psi\in {\cal H}$ 
\be \label{4.15}
||\psi_{\phi,S_n}-\psi||\le \int dt \phi(t) 
||[\pi(W_t(S_n))-\mbox{id}_{{\cal H}}]\psi|| 
\ee
can be made arbitrarily small due to the assumed weak continuity of the 
Weyl elements by suitably restricting the support of $\phi$. Hence 
the $[\pi(T_s)1^\nu]_{\phi,S_n}$ lie dense as $s,\nu,\phi$ vary for any
$S_n$. Thus
\ba \label{4.16}
&& ||\pi(T_s)1^\nu_{\phi,S_n}-[\pi(T_s)1^\nu]_{\phi,S_n}||
\nonumber\\
&=&
||\int dt \phi(t) [\pi(T_s)\pi(W_t(S_n))-\pi(W_t(S_n))\pi(T_s)]1^\nu ||
\nonumber\\
&\le&
\int dt \phi(t) ||[\pi(T_s)\pi(W_t(S_n))-\pi(W_t(S_n))\pi(T_s)]1^\nu ||
\nonumber\\
&=& 
\int dt \phi(t) 
||[\pi(W_t(S_n))^{-1}\pi(T_s)\pi(W_t(S_n))-\pi(T_s)]1^\nu||
\nonumber\\
&=& 
\int dt \phi(t) 
||\pi(W_t(S_n)^{-1} T_s W_t(S_n)-T_s)1^\nu||
\nonumber\\
&=& 
\int dt \phi(t) 
||W_t(S_n)^{-1} T_s W_t(S_n))-T_s||_\nu
\nonumber\\
&\le& 
\int dt \phi(t) 
||W_t(S_n)^{-1} T_s W_t(S_n))-T_s||_\infty
\ea
where in the third step we have used unitarity of the Weyl elements,
in the fourth we have used the representation property, in the fifth 
we have used that the argument of $\pi$ is just a function which acts
as a multiplication operator and in the sixth step we have again used 
continuity with respect to the sup norm $||.||_\nu\le ||.||_\infty$. 

Since $T_s$ is a continuous (on $\ab$) cylindrical function over 
$\gamma(s)$ 
and right translation $A(e)\mapsto e^{t n^j\tau_j} A(e)$ is continuous in
$t$, it follows that (\ref{4.16}) can be made arbitrarily small by 
suitably restricting the support of $\phi$.\\
$\Box$\\
\begin{Corollary} \label{col4.1} ~~~~\\
The set of vectors $\pi(f)1^\nu_{\phi,S_n}$, as $f\in \mbox{Cyl}^\infty, 
\nu$ vary, form a dense set of $C^\infty$ vectors for the self-adjoint
generator $\pi(E(S_n))$ of $\pi(W_t(S_n))$, more precisely
\be \label{4.17}
-i\pi(E(S_n))\pi(f)1^\nu_{\phi,S_n}:=
[\frac{d}{dt}]_{t=0} \pi(W_t(S_n))\pi(f) 1^\nu_{\phi,S_n}
=\pi(Y(S_n)f) 1^\nu_{\phi,S_n}+\pi(f) 1^\nu_{-\dot{\phi},S_n}
\ee
In particular, ${\cal D}(S_n)$ is a dense invariant domain for 
$\pi(E(S_n))$.
\end{Corollary}
Proof of corollary \ref{col4.1}:\\
We have 
\ba \label{4.18}
\pi(W_t(S_n))\pi(f) 1^\nu_{\phi,S_n}
&=&\pi(W_t(S_n) f W_t(S_n)^{-1})
\int ds \phi(s) \pi(W_{t+s}(S_n)) 1^\nu
\nonumber\\
&=& \pi(W_t(S_n) f W_t(S_n)^{-1})
\int ds \phi(s-t) \pi(W_s(S_n)) 1^\nu
\ea
Observing the Weyl relations, differentiation of (\ref{4.18}) in the 
strong sense yields 
(\ref{4.17}) (for details see \cite{23}).\\
$\Box$\\
\\
{\bf Step 2:}\\
\begin{Lemma} \label{la4.2} ~~~\\
Let $S$ be an analytic, oriented, open surface and $e$ an oriented, 
analytic path. Let $p_1,..,p_m,\;m\ge 1$ be fixed interior points of $e$ 
and choose $\sigma_k\in \{-1,+1\},\;k=1,..,m$. Then there exists an 
analytic diffeomorphism $\varphi_{m,\vec{\sigma}}$ such 
that $\varphi_{m,\vec{\sigma}}^{-1}(S)$ intersects $e$ precisely in the 
points $p_1,..,p_m$ and such that
\be \label{4.19} 
\sigma([\varphi_{m,\vec{\sigma}}^{-1}(S)]_{p_k},[e]_{p_k})=\sigma_k
\ee
\end{Lemma}
Proof of lemma \ref{la4.2}:\\
We certainly find a smooth diffeomorphism 
$\varphi^\infty_{m,\vec{\sigma}}$ with the required properties. 
Consider
a compact set $C$ containing $e$ and the algebra of real valued,
continuous functions $S_C$ generated by the functions 
\be \label{4.20}
\{\varphi^a_{|C};\;\varphi\in 
\mbox{Diff}^\omega(\sigma),\;a=1,..,D\}
\ee
Then $S_C$ separates the points of $C$ (choose $\varphi=\mbox{id}_\sigma$)
and does not leave any point $x_0\in C$ invariant.
Hence by the Weierstrass theorem
$S_C$ is dense in the set $C(C)$ of continuous functions on $C$ and since
$\varphi^\infty_{m,\vec{\sigma}}$ is continuous we find an analytic 
diffeomorphism $\varphi^0_{m,\vec{\sigma}}$ that approximates it uniformly
on $C$ in the sup norm. While the intersection points $p'_k$ 
of $[\varphi^0_{m,\vec{\sigma}}]^{-1}(S)$ with $e$ may not yet
coincide with the $p_k$ (although they are arbitrarily close), it is 
nevertheless true that 
$\sigma([(\varphi^0_{m,\vec{\sigma}})^{-1}(S)]_{p'_k},[e]_{p'_k})=\sigma_k$ 
since the $\sigma$ functions take only discrete values.

We will now construct successively analytic diffeomorphisms 
$\varphi_k,\;k=1,..,m$ which preserve $e$ such that 
$\varphi_k^{-1}(p_l)=p_l,\;l=0,..,k-1,\;\varphi_k^{-1}(p'_k)=p_k,\;  
\varphi_k^{-1}(p_{m+1})=p_{m+1}$ where $p_0=b(e),\;p_{m+1}=f(e)$
and such that the $\sigma_k$ are not changed. Then 
\be \label{4.21}
\varphi_{m,\vec{\sigma}}:=\varphi^0_{m,\vec{\sigma}}\circ
\varphi_1\circ ..\circ \varphi_m
\ee
provides the searched for diffeomorphism.

To construct $\varphi_k$ explicitly, choose w.l.g. an analytic coordinate 
system such that $e$ coincides with the interval $[0,1]$ of the 
$x^1-$axis (if $e$ does not lie entirely within the domain of a chart,
replace $e$ by a closed segment of it that does in what follows). Then 
$p_k=(x_k,0,..,0)$ and $p'_k(y_k,0,..,0)$ are 
the coordinates of the points in question and we label them in such a way
that $x_0=0<x_1<..<x_m<x_{m+1}=1,\; 0<y_1<..<y_m<1$. 
The situation for $\varphi_k$
is such that $y_l=x_l,\;l=1,..,k-1$ already while the $y_l,\;
l=k,..,m$ are unspecified. Thus,
the idea is to construct an 
analytic vector field $x\mapsto v_k(x)$ on $\Rl$ which has zeroes at the 
points $x_0=0, x_1,..,x_{k-1}, x_{m+1}=1$ and whose flow maps $y_k$ to 
$x_k$. Consider the analytic vector field on $\Man$ defined in our 
coordinate system by 
$\vec{v}_k(\vec{x})=(v_k(x^1),0,..,0)$. The integral curves 
$c^{\vec{v}_k}_{\vec{x}}(t)$ it generates defines a one parameter family 
of analytic diffeomorphisms 
$\varphi^{\vec{v}_k}_t(\vec{x}):=c^{\vec{v}_k}_{\vec{x}}(t)$ of the 
form 
\be \label{4.22}
\varphi^{\vec{v}_k}_t(\vec{x})=(\varphi^{v_k}_t(x^1),x^2,..,x^D)
\ee
where $\varphi^{v_k}_t(x^1)$ is the corresponding one parameter group
in $\Rl$.
It follows that 
$\varphi^{\vec{v}_k}_t(\vec{x})$ for any $t$ is 
just an $x^1$ 
dependent translation along the $x^1-$axis so that the $\sigma_k$ are 
left invariant. It remains to construct $v_k(x)$ and to choose $t$ with 
the required properties. Notice that if $v_k$ has a zero at some $x$
then $x$ is a fixed point of $\varphi^{v_k}_t$.

Our ansatz is given for $\delta>0$ by 
\be \label{4.23}
v_k(x)=\mbox{sgn}(x_k-y_k) 
[1-e^{-\frac{(x-1)^2}{2\delta^2}}]
\prod_{l=0}^{k-1} 
[1-e^{-\frac{(x-x_l)^2}{2\delta^2}}]
\ee
where sgn$(x)=1,-1,0$ if $x>,<,=0$ is the sign function. The flow thus by 
construction preserves $x=0,x_1,..,x_{k-1},1$ and moves, apart from the 
fix points, into positive
or negative $x^1-$direction respectively if $x_k-y_k>/<0$ respectively
(if $x_k=y_k$ already we can choose $v_k=0$ obviously). 
We set $\alpha=\mbox{sgn}(x_k-y_k)$ in the remainder of this proof.

Consider the function $f(t,x):=v_k(x)$ which does not depend explicitly
on $t$. We have 
\be \label{4.24}
|f(t,x)-f(t,x')| = |\int_x^{x'} dy v'_k(y)| \le 
\int_x^{x'} dy |v'_k(y)| \le \frac{m+2}{\delta} |x-x'|
\ee
where we have used that for the function $g(x)=1-e^{-x^2/(2\delta^2)}$
holds $0\le g(x)\le 1$ and $|g'(x)|\le e^{-1/2}/\delta$. Thus the 
function $f(t,x)$ satisfies a global Lipschitz condition in the open
domain $G=\Rl\times [a,b]$ where $[0,1]\subset [a,b]$ defines the 
boundaries of our chart in $x^1$ direction. By the 
existence and 
uniqueness theorem of Picard -- Lindel\"of for the differential equation 
$\dot{x}(t)=f(t,x)$, for any $(t_0,x_0)\in G$ there exists $\epsilon>0$
and a unique solution in $t\in [t-\epsilon,t+\epsilon]$ with initial
condition $x(t_0)=x_0$. The number $\epsilon$ is bounded by 
$\min(r,\frac{r}{c})$ where $r$ is the maximal number such that 
$V_r:=(t_0-r,t_0+r)\times (x_0-r,x_0+r)$ is still in $G$ while 
$c=\sup_{(t,x)\in V_r} |f(t,x)|$. In our case $t_0=0,\;x_0=y_k$, $r=
\min(y_k-a,b-y_k),\;c=1$ so 
that 
$\epsilon=r$ is arbitrarily large by choosing coordinates in 
which $|a|,|b|$ become arbitrarily large and the segment of $e$ 
considered remains the image of $[0,1]$. It 
follows that 
we find a solution of the differential equation 
$\dot{c}_{y_k}(t)=v_k(c_{y_k}(t))$ with $c_{y_k}(0)=y_k$ which exists for 
$t\in [-r,r]$ for arbitrarily large but finite $r$.  

Notice that after applying the $k-1$th diffeomorphism $\varphi_{k-1}$ 
we have already achieved that $y_l=x_l,\;l=1,..,k-1$ (the order of the 
points $y_k$ cannot be changed by a diffeomorphism) and 
our job is now 
to construct $\varphi_k$ which has to move $y_k>y_{k-1}=x_{k-1}$ to 
$x_k$ while leaving $x=0,x_1,..,x_{k-1},1$ fixed. 
Define now 
\be \label{4.25}
\mu:=\frac{1}{2}\min(y_k-x_{k-1},\; x_k-x_{k-1},\;1-\frac{1+\alpha}{2} x_k
-\frac{1-\alpha}{2} y_k\}
\ee
and the interval $I_k=[x_{k-1}+\mu,1-\mu]$. Consider the differential
equation $\dot{x}(t)=v_k(x(t)),\;x(0)=y_k$. As long as 
$x(t)\in I_k$ we have $|x(t)-1|\ge\mu,\;|x(t)-x_l|\ge \mu;\;l=0,..,k-1$
and thus
$\alpha v_k(x(t))\ge[1-\exp(-\frac{\mu^2}{2\delta^2})]^{k+1}$. On the 
other hand,
$\alpha v_k(x(t))\le 1$ for all $t\in \Rl$. From the integral equation
\be \label{4.26}
x(t)=y_k+\int_0^t ds v(x(s))
\ee
we thus conclude for $t\ge 0$
\be \label{4.27}
y_k+t\ge x(t)\ge y_k+[1-\exp(-\frac{\mu^2}{2\delta^2})]^{k+1} t
\ee
for $\alpha=1$ and inequality signs reversed in case $\alpha=-1$,
as long as $x(t)\in I_k$. 
Now $y_k-(x_{k-1}+\mu)\ge\frac{y_k-x_{k-1}}{2}\ge 0$ and 
$1-\mu-y_k\ge \frac{1+x_k}{2}-y_k\ge 0$ if $\alpha=1$ because $x_k,1\ge 
y_k$ or
$1-\mu-y_k\ge \frac{1+y_k}{2}-y_k\ge 0$ if $\alpha=-1$. Hence
$y_k\in I_k$ and since all tree terms in (\ref{4.27}) are monotonously
increasing (decreasing) with $t$ for $\alpha=1$ ($\alpha=-1$) we can 
guarantee $x(t)\in I_k$ be requiring 
$t\le T$ where $y_k+T=1-\mu$ for $\alpha>0$ and 
$y_k-T=x_{k-1}+\mu$ for $\alpha<0$.
We conclude that 
\be \label{4.28a}
x(T)\ge y_k+[1-\exp(-\frac{\mu^2}{2\delta^2})]^{k+1} (1-\mu-y_k)
\ee
for $\alpha>0$ and 
\be \label{4.28b}
x(T)\le y_k+[1-\exp(-\frac{\mu^2}{2\delta^2})]^{k+1} (y_k-\mu-x_{k-1})
\ee
for $\alpha<0$.
We claim that we can choose $\delta$ small enough such that the right 
hand side of (\ref{4.28a}) ((\ref{4.28b})) is bigger (lower) than $x_k$. Thus 
we must satisfy
\be \label{4.29a}
\ln(1-e^{-\frac{\mu^2}{2\delta^2}})\ge \frac{1}{k+1}
\ln(\frac{x_k-y_k}{1-\mu-y_k})
\ee
for $\alpha>0$ and 
\be \label{4.29b}
\ln(1-e^{-\frac{\mu^2}{2\delta^2}})\ge \frac{1}{k+1}
\ln(\frac{y_k-x_k}{y_k-\mu-x_{k-1}})
\ee
for $\alpha<0$.
Notice that the argument of the logarithm on the right hand side of 
(\ref{4.29a}) and (\ref{4.29b}) respectively is smaller than one since 
$x_k<1-\mu$ and $\mu+x_{k-1}<x_k$ respectively. Since both 
hand sides of (\ref{4.29a}), (\ref{4.29b}) are negative we must show
\be \label{4.30}
|\ln(1-e^{-\frac{\mu^2}{2\delta^2}})|\le \frac{1}{k+1}
|\ln(\frac{x_k-y_k}{1-\mu-y_k})|
\ee
for $\alpha>0$ and similar for $\alpha<0$.
Let $h(x)=-\ln(1-x),\;0\le x<1$. We claim that $h(x)\le\sqrt{x}$.
For $k(x):=\sqrt{x}+\ln(1-x)$ we find $k(0)=0,\;k'(x)>0$ for 
$x<(\sqrt{2}-1)^2$. Hence, (\ref{4.30}) holds provided that
\be \label{4.31}
e^{-\frac{\mu^2}{4\delta^2}}\le \frac{1}{k+1}
|\ln(\frac{x_k-y_k}{1-\mu-y_k})|
\ee
holds and $\delta$ is small enough so that 
$e^{-\frac{\mu^2}{2\delta^2}}<3-\sqrt{8}<1$.
We conclude
\be \label{4.32}
-\frac{\mu^2}{4\delta^2}\le \ln(\frac{1}{k+1}
|\ln(\frac{x_k-y_k}{1-\mu-y_k})|)
\ee
which is trivially satisfied if the right hand side is positive 
(which it will not be when $k$ is large). If the right hand side is 
negative we find 
\be \label{4.33}
\delta^2<\min(\frac{\mu^2}{2\ln(3-\sqrt{8})},
\frac{\mu^2}{4|\ln(\frac{1}{k+1}|\ln(\frac{x_k-y_k}{1-\mu-y_k})|)|})
\ee
which can always be satisfied. The case $\alpha<0$ is similar.

From the continuity of the solution $x(t)$ we conclude that there exists 
$t_k\in [0,T]$ such that $x(t)=x_k$. Thus, if 
$t\mapsto \varphi_{k,t}$ is the one parameter family of 
diffeomorphisms generated by $-v_k$ we define $\varphi_k:=\varphi_{k,t_k}$
and have $\varphi_k(x_l)=x_l,\;l=0,..,k-1,\;l=m$ and 
$\varphi_k(y_k)=x_k$ as desired, which concludes our induction step.\\
$\Box$\\
\\
{\bf Step 3:}\\
This step contains the main argument in our proof. The self-adjoint 
generator $\pi(Y_n(S))$ is symmetric and has dense domain 
${\cal D}(S_n)$ for any $n$. From (\ref{4.17}) we find the symmetry 
condition
\ba \label{4.41}
&&<\pi(f)1^\nu_{\phi,S_n},\pi(E(S_n))\pi(f')1^{\nu'}_{\phi',S_n}>
\nonumber\\
&=&i<\pi(f)1^\nu_{\phi,S_n},
\pi(Y(S_n)f') 1^{\nu'}_{\phi',S_n}+\pi(f') 1^{\nu'}_{-\dot{\phi}',S_n}>
\nonumber\\
&=&
<\pi(E(S_n))\pi(f)1^\nu_{\phi,S_n},\pi(f')1^{\nu'}_{\phi',S_n}>
\nonumber\\
&=&-i<\pi(Y(S_n)f) 1^\nu_{\phi,S_n}+\pi(f) 1^\nu_{-\dot{\phi},S_n},
\pi(f')1^{\nu'}_{\phi',S_n}>
\ea
Choose $f=1$ then we obtain the master condition
\ba \label{4.42}
&&-<1^\nu_{\phi,S_n},\pi(Y(S_n)f') 1^{\nu'}_{\phi',S_n}>
\nonumber\\
&=& <1^\nu_{\phi,S_n},\pi(f') 1^{\nu'}_{-\dot{\phi}',S_n}>
+ <1^\nu_{-\dot{\phi},S_n},\pi(f') 1^{\nu'}_{\phi',S_n}>
\ea
We now split the proof into the Abelian case and the non-Abelian case 
because we are able to use less assumptions in the Abelian case. 
We will also indicate what goes wrong when trying to repeat the Abelian 
proof for the non-Abelian case which might might lead the ambitious reader
to a method for how to circumvent the obstacle. Also, we offer two 
different proof methods in the non-Abelian case.\\
\\
{\bf Abelian Case}\\
We will carry out the proof just for one copy of $U(1)$, the general case 
is similar. In that case $n_j=1$ and we can drop the label $n$ from 
$Y_n(S)$ and $W^n_t(S)$. 
We employ some of the ideas already used in \cite{10}.\\
Choose $f'=\alpha_{\phi^{e,S}_m}(T_s)$ where $T_s$ is a spin-network 
function and $\varphi^{e,S}_m$ is an analytic diffeomorphism such that 
$(\varphi^{e,S}_m)^{-1}(\gamma(s))$ intersects $S$ in precisely $m$
distinct points $p_1,..,p_m$ which are interior points of the edge $e\in 
E(\gamma(s))$ and such that 
$\sigma([S]_{p_k},[(\varphi^{e,S}_m)^{-1}(e)]_{p_k})=1$. The existence of 
such a diffeomorphism can be deduced from the first half of 
lemma \ref{la4.2}. Notice that we only need one of the $2^m$ 
diffeomorphisms constructed in lemma \ref{la4.2} for the Abelian case.
The irreducible representations of $U(1)$ are one dimensional so that
$\pi_e$ is just determined by an integer $\lambda_e$. Moreover 
$Y(S)f'=m\lambda_e f'$ is an eigenfunction. Thus (\ref{4.42}) becomes
\be \label{4.44}
-m\lambda_e  <1^\nu_{\phi,S},\pi(f') 1^{\nu'}_{\phi',S}>
=<1^\nu_{\phi,S},\pi(f') 1^{\nu'}_{-\dot{\phi}',S}>
+ <1^\nu_{-\dot{\phi},S},\pi(f') 1^{\nu'}_{\phi',S}>
\ee
We now estimate the left hand side and the right hand side of 
(\ref{4.44}). For the right hand side we have 
\ba \label{4.45}
&&
|<1^\nu_{\phi,S},\pi(f') 1^{\nu'}_{-\dot{\phi}',S}>
+ <1^\nu_{-\dot{\phi},S},\pi(f') 1^{\nu'}_{\phi',S}>|
\nonumber\\
&\le& 
||1^\nu_{\phi,S}||\;||\pi(f') 1^{\nu'}_{-\dot{\phi}',S}||+
||1^\nu_{-\dot{\phi},S}||\;||\pi(f') 1^{\nu'}_{\phi',S}||
\nonumber\\&\le& 
||f'||_\infty[ ||1^\nu_{\phi,S}||\;||1^{\nu'}_{-\dot{\phi}',S}||+
||1^\nu_{-\dot{\phi},S}||\;||1^{\nu'}_{\phi',S}||]
\nonumber\\
&\le& 
||T_s||_\infty[||\phi||_1\;||\dot{\phi}'||_1+
||\dot{\phi}||_1\;||\phi'||_1]
\ea
where in the second step we have used the Cauchy-Schwarz inequality,
in the third the elementary continuity of the operator norm 
with respect to the $C^\ast-$norm which is diffeomorphism invariant 
(so the dependence on $\varphi^{e,S}_m$ drops out) and 
finally 
$||\psi_\phi||\le \int dt |\phi(t)|\;\;\; ||\pi(W_t(S))\psi||\le 
||\phi||_1\;||\psi||$ due to unitarity. For the left hand side 
of (\ref{4.44}) we have
\ba \label{4.46}
&& |-m\lambda_e  <1^\nu_{\phi,S},\pi(f') 1^{\nu'}_{\phi',S}>|
\nonumber\\
&\ge& 
m |\lambda_e|  
[|<1^\nu,\pi(f') 1^{\nu'}>|
-|<1^\nu_{\phi,S}-1^\nu,\pi(f') 1^{\nu'}>|
-|<1^\nu_{\phi,S},\pi(f') (1^{\nu'}_{\phi',S}-1^{\nu'})>|]
\nonumber\\ &\ge& 
m |\lambda_e|  
[|<1^\nu,\pi(f') 1^{\nu'}>|
-||T_s||_\infty\{ 
||1^\nu_{\phi,S}-1^\nu||+||1^{\nu'}_{\phi,S}-1^{\nu'}||\}]
\ea
Suppose now that for $\lambda_e\not=0$ we have  
\be \label{4.47}
\delta=|<1^\nu,\pi(f') 1^{\nu'}>|=
|<1^\nu,\pi(T_s) 1^{\nu'}>|>0
\ee
where we have made use of diffeomorphism invariance of $1^\nu,1^{\nu'}$
so that the only dependence on the intersection number $m$ in (\ref{4.46})
is just the prefactor. Due to weak continuity of the fluxes, we 
can restrict the support of $\phi,\phi'$ to 
$|t|\le \epsilon(\nu,\delta/4,S,s)$ and 
$|t|\le \epsilon(\nu',\delta/4,S,s)$ respectively such that
$||1^\nu_{\phi,S}-1^\nu||\le \delta/(4||T_s||_\infty||)$ 
and $||1^{\nu'}_{\phi,S}-1^{\nu'}||\le \delta/(4||T_s||_\infty||)$ 
respectively.
Now choose for the so chosen $\phi,\phi'$ the intersection number
$$
m>2\frac{||f'||_\infty[||\phi||_1\;||\dot{\phi}'||_1
+||\dot{\phi}||_1\;||\phi'||_1]}{\delta}
$$
Then we have produced a contradiction, thus 
$<1^\nu,\pi(T_s) 1^{\nu'}>=0$ unless $\lambda_e=0$. Since $e\in 
E(\gamma(s))$ was arbitrary and since 
$1^\nu,\pi(f) 1^{\nu'}>=\mu_\nu(f) \delta_{\mu\nu}$ we conclude 
\be \label{4.48}
\mu_\nu(T_s)= \left\{ \begin{array}{ll}
1 & \mbox{ : $s$ trivial} \\
0 & \mbox{ : otherwise} 
\end{array} \right.
\ee
which is one of the equivalent definitions of the Ashtekar-Lewandowski
measure.\\
\\
{\bf Non-Abelian Case}\\
Before we proceed to the proof under the additional assumption iv)
let us first outline why the strategy for the Abelian case does 
not work here. The basic problem is that no $f'=T_s$  
spin-network function is an eigenfunction $Y_n(S) T_s\not\propto T_s$
of the fluxes. Without this being the case, we have no chance to 
get something like $Y_n(S)\alpha_{\varphi^{e,S}_m}(T_s)\propto m$
which was crucial in the Abelian case. Thus, in order to get this 
proportionality the idea is to  consider something squares of the form
$Y_n(S)^2\alpha_{\varphi^{e,S}_m}(T_s)$ which have better chances because 
if we take suitable linear combinations then we obtain Laplacians which do 
have the $T_s$ as eigenfunctions. 

In order to get such squares we just have to iterate (\ref{4.42}) or 
(\ref{4.43}) by choosing $f'=Y_n(S)f$ or $f'=Y^+_j(S)f$ for some $f$
to be suitably chosen. This results in  
\ba \label{4.49}
&&<1^\nu_{\phi,S_n},\pi(Y(S_n)^2f) 1^{\nu'}_{\phi',S_n}>
\nonumber\\
&=& 
<1^\nu_{\phi,S_n},\pi(f) 1^{\nu'}_{\ddot{\phi}',S_n}>
+2<1^\nu_{-\dot{\phi},S_n},\pi(f) 1^{\nu'}_{-\dot{\phi}',S_n}>
+ <1^\nu_{\ddot{\phi},S_n},\pi(f) 1^{\nu'}_{\phi',S_n}>
\ea
Let us now choose $f=\alpha_{\varphi^{e,S}_{m,\sigma}}^{-1}(T_s)$ for any 
spin network
function $T_s$ where $\varphi^{e,S}_{m,\sigma}$ is the diffeomorphism 
constructed in lemma \ref{la4.2} such that 
$\varphi^{e,S}_{m,\sigma}(S)$ intersects $\gamma(s)$ precisely in
$m$ interior points $p_k$ of $e\in E(\gamma(s))$ with relative 
orientation $\sigma_k$. Let us write 
$e=f_1^{-1}\circ e_1\circ f_2^{-1}\circ e_2\circ..\circ f_m^{-1}\circ e_m$
where $p_k=f_k\cap e_k$. Then 
\ba \label{4.51}
&&\alpha_{\varphi^{e,S}_{m,\sigma}}(Y_n(S)^2) T_s
\nonumber\\
&=& 
\sum_{I,J=1}^m \sigma_I\sigma_J 
(R^n_{e_I}-R^n_{f_I})(R^n_{e_J}-R^n_{f_J})T_s
\nonumber\\
&=& 
\sum_{I=1}^m  
[2 (R^n_{e_I})^2+2(R^n_{f_I})^2]T_s
+8\sum_{I<J}^m \sigma_I\sigma_J 
R^n_{e_I} R^n_{e_J}T_s
\ea
where in the second step we have used gauge invariance of $T_s$ at $p_k$,
that is, $(R^n_{e_I}+R^n_{f_I})T_s=0$. Now the idea would be to 
choose $n^k=\delta_{jk}$, to sum over $j$ and to average over 
the $2^m$ possible choices for $\sigma=(\sigma_1,..,\sigma_m)$. Thus
\be \label{4.52}
2^{-m}\sum_j \sum_\sigma 
\alpha_{\varphi^{e,S}_{m,\sigma}}(Y_j(S)^2) T_s
=-8m \lambda_{\pi_e} T_s
\ee
where $\pi_e=\pi_e(s)$ is the irreducible representation of $G$ labeling 
$e$ and $-\lambda_{\pi_e}\le 0$ is the corresponding eigenvalue of the
Laplacian. In 
order to exploit this, we write (\ref{4.49}) as
\ba \label{4.53}
&&
2^{-m}\sum_j \sum_\sigma 
<1^\nu_{\phi,S_j},\pi(Y(S_j)^2 
\alpha_{\varphi^{e,S}_{m,\sigma}}^{-1}(T_s)) 1^{\nu'}_{\phi',S_j}>
\nonumber\\
&=& 
2^{-m}\sum_j \sum_\sigma 
<1^\nu_{\phi,(\varphi^{e,S}_{m,\sigma})^{-1}(S_j)},
\pi(\alpha_{\varphi^{e,S}_{m,\sigma}}(Y_j(S)^2)T_s) 
1^{\nu'}_{\phi',(\varphi^{e,S}_{m,\sigma})^{-1}(S_j)}>
\nonumber\\
&=& 
2^{-m}\sum_j \sum_\sigma 
[<1^\nu_{\phi,(\varphi^{e,S}_{m,\sigma})^{-1}(S_j)},
\pi(T_s) 1^{\nu'}_{\ddot{\phi}',(\varphi^{e,S}_{m,\sigma})^{-1}(S_j)}>
+2<1^\nu_{-\dot{\phi},(\varphi^{e,S}_{m,\sigma})^{-1}(S_j)},
\pi(T_s) 1^{\nu'}_{-\dot{\phi}',(\varphi^{e,S}_{m,\sigma})^{-1}(S_j)}>
\nonumber\\
&& + <1^\nu_{\ddot{\phi},(\varphi^{e,S}_{m,\sigma})^{-1}(S_j)},
\pi(T_s) 1^{\nu'}_{\phi',(\varphi^{e,S}_{m,\sigma})^{-1}(S_j)}>]
\ea
Now the trouble is that a $(j,\sigma)-$dependence has entered the states
$1^\nu_{\phi,(\varphi^{e,S}_{m,\sigma})^{-1}(S_j)}$ which prevents us from 
using (\ref{4.52}). In order to get rid of the $(j,\sigma)-$dependence
of the states, we would need to construct states which are common 
$C^\infty-$vectors for all the $\alpha_{\varphi^{e,S}_{m,\sigma}}(Y_j(S))$.
That is certainly possible: First of all we pass to the $Y^+_j(S)$ 
of appendix \ref{sb}
and choose as $S$ a surface which is the disjoint union of $m$ pieces 
$S_I$. Then define
\be \label{4.54}
1^\nu_{\phi,S}:=
\int_G d\mu_H(g_1)..
\int_G d\mu_H(g_m) \phi(g_1,..,g_m) \prod_{I=1}^m \pi(W_{g_I}(S_I)) 1^\nu
\ee
and if we can arrange that 
$S_\sigma=\varphi^{e,S}_{m,\sigma})^{-1}(S)=\cup_I \sigma_I S_I$
then 
\be \label{4.55}
Y^+_j(S_\sigma) 1^\nu_{\phi,S}=1^\nu_{-\sum_I \sigma_I R^j_I \phi,S}
\ee
for all $j,\sigma$.
However, the sum over $I$ involved in (\ref{4.55}) destroys our estimates 
performed in the Abelian case for which it was crucial that the right hand 
side of (\ref{4.44}) was already independent of $m$. This is the obstacle 
that prevents us from using the idea employed for the Abelian case.
Thus, in order to proceed, let us make the additional assumption 
iv). While we feel that this is not necessary, we could so far not find 
a way to circumvent the obstacle just mentioned.\\
\\
According to assumption iv) we can take the limit 
$\phi(t)\to\delta(t)$ in (\ref{4.53}). Then the $(j,\sigma)-dependence$
disappears from the states,
$1^\nu_{\phi,(\varphi^{e,S}_{m,\sigma})^{-1}(S_j)}\to 1^\nu$, and we find
\ba \label{4.56}
&&
2^{-m}\sum_j \sum_\sigma 
<1^\nu,\pi(Y(S_j)^2 
\alpha_{\varphi^{e,S}_{m,\sigma}}^{-1}(T_s)) 1^{\nu'}>
\nonumber\\
&=&- 8\lambda_{\pi_e} m <1^\nu, \pi(T_s) 1^{\nu'}>
\nonumber\\
&=&
\sum_j 
[<1^\nu,\pi(2^{-m}\sum_\sigma \alpha_{\varphi^{S,e}_{m,\sigma}}^{-1}(T_s))
\pi(Y_j(S)^2)1^{\nu'}>
+2<\pi(Y_j(S))1^\nu,\pi(2^{-m}\sum_\sigma 
\alpha_{\varphi^{S,e}_{m,\sigma}}^{-1}(T_s))\pi(Y_j(S))1^{\nu'}>
\nonumber\\
&& +<\pi(Y_j(S)^2)1^\nu,\pi(2^{-m}\sum_\sigma 
\alpha_{\varphi^{S,e}_{m,\sigma}}^{-1}(T_s))1^{\nu'}>]
\ea
Estimating the right hand side of (\ref{4.57}) from above we have 
\ba \label{4.57}
&& 8\lambda_{\pi_e} m |<1^\nu, \pi(T_s) 1^{\nu'}>|
\nonumber\\
&\le& ||T_s||_\infty
\sum_j 
[||\pi(Y_j(S)^2)1^{\nu'}||
+||\pi(Y_j(S))1^\nu||\;||\pi(Y_j(S))1^{\nu'}||
+||\pi(Y_j(S)^2)1^\nu||]
\ea
Due to the diffeomorphism invariance of $1^\nu$ the right hand side of 
(\ref{4.57}) no longer depends on $m$ in contrast to the left hand side 
which implies as in the Abelian case that $\mu_\nu=\mu_0$ is the 
Ashtekar Lewandowski measure.\\
\\
{\bf Step 4:}\\
We stress that the additional requirement iv) is only necessary 
in step 3 of the proof. The remainder is again independent of that.
It rests crucially on our already available knowledge that all $\mu_\nu$
equal the Ashtekar-Lewandowski measure. 

We make the general ansatz 
\be \label{4.60}
\pi(W_t(S_n))=[M_t(S_n)]\;[W_t(S_n)\otimes\pi(1)]
\ee
where the second factor acts diagonally in each entry $\psi_\nu$ of 
$\psi=\oplus \psi_\nu$ by left translation on $\ab$ and $M_t(S_n)$ is an 
operator valued matrix of the form 
\be \label{4.61}
M_t(S_n)\psi=\sum_{\nu,\nu'} ([M_t(S_n)]_{\nu\nu'}\cdot \psi_{\nu'}) 1^\nu
\ee
Using the representation property 
\be \label{4.62}
\pi(W_t(S_n))\pi(f)\pi(W_t(S_n))^{-1}
=\pi(W_t(S_n) f W_t(S_n)^{-1})
\ee
and that $f$ is arbitrary we conclude that 
\be \label{4.63}
{[}(M_t(S_n))_{\nu\nu'},f]=0
\ee
for any $f\in \mbox{Cyl}_b$ and any $\nu,\nu'$. It follows that 
$M_t(S_n)$ is a multiplication operator valued matrix.

Due to the unitarity of $\pi(W_t(S_n))$ we compute 
\ba \label{4.64}
1 &=& ||\pi(W_t(S_n))1^\nu||^2=||M_t(S_n) 1^\nu||^2=
||\sum_{\nu'} ([M_t(S_n)]_{\nu'\nu}\cdot 1) 1^{\nu'}||^2
\nonumber\\
&=&
\sum_{\nu'} ||[M_t(S_n)]_{\nu'\nu}||^2_{\mu_{\nu'}}
\ea
for any $\nu$. Thus all the matrix entries $[M_t(S_n)]_{\nu\nu'}$ are 
$L_2(\ab,d\mu_0)$ functions and we can expand them in terms of 
spin network functions
\be \label{4.65}
{[}M_t(S_n)]_{\nu\nu'}=\sum_s [Z_t(S_n,s)]_{\nu\nu'} T_s
\ee
for some complex valued coefficients $[M_t(S_n,s)]_{\nu\nu'}$ 
of which all but countably many must vanish. 

From diffeomorphism covariance (remember that $U_\pi$ is the natural 
representation of the diffeomorphism group)
we have for $n_j(x)=n_j=const.$
\ba \label{4.67}
U_\pi(\varphi)\pi(W^n_t(S))U_\pi(\varphi)^{-1}
&=& \pi(\alpha_\varphi(W^n_t(S)))=\pi(W^n_t(\varphi^{-1}(S))
=M^n_t(\varphi^{-1}(S))[W^n_t(\varphi^{-1}(S))\otimes\pi(1)]
\nonumber\\
&=&
[U_\pi(\varphi)M_t(S_n)U_\pi(\varphi)^{-1}]\;
[U_\pi(\varphi)[W^n_t(S)\otimes\pi(1)]U_\pi(\varphi)^{-1}]
\nonumber\\
&=& [U_\pi(\varphi)M_t(S_n)U_\pi(\varphi)^{-1}]\;
[W^n_t(\varphi^{-1}(S))\otimes\pi(1)]
\ea
and 
\be \label{4.68}
U_\pi(\varphi) \pi(T_s) U_\pi(\varphi^{-1})=\pi(\alpha_\varphi(T_s))
=\pi(T_{\varphi(s)})
\ee
where 
$\varphi(s)=(\varphi^{-1}(\gamma(s)),\vec{\pi(s)},\vec{M}(s),\vec{N}(s))$,
we conclude by comparing coefficients that
\be \label{4.69}
[Z^n_t(\varphi^{-1}(S),s)]_{\nu,\nu'}=[Z^n_t(S,\varphi(s))]_{\nu\nu'}
\ee
for any $s$.

Suppose now that $\gamma(s)\not=\emptyset$ for $D>3$ or 
$\gamma(s)\not=\emptyset,\partial S$ for $D=3$  (neither the empty 
graph nor the graph formed by the boundary of the closure of $S$, that 
is $\gamma(s)\not=\overline{S}-\mbox{Int}(S)$)
or $\gamma(s)\not=\emptyset,\overline{S}$ for $D=2$.
Then we find a countably infinite number of analytic diffeomorphisms
$\varphi_k$ which leave $S$ invariant but such that the 
$\varphi_k(\gamma(s))$ are mutually different. To construct such a 
diffeomorphism for $D>1$, simply take any analytical vector field which 
is everywhere tangent to $S$ and tangent to $\partial S$ (e.g.
vanishes on (non differentiable points of) $\partial S$). Then
$S,\partial S$ are left invariant as sets, but not pointwise, by the 
one parameter group of analytical diffeomorphisms generated by that 
vector field. Thus for $D>3$ even a graph which lies completely within 
the closure $\overline{S}$ can be mapped non-trivially, 
for $D=2$ the graph cannot be mapped non-trivially only if $\gamma(s)=S$
and for $D=3$ we must have $\gamma(s)=\partial S$. 
Thus, unless one of the cases indicated holds, we always 
find a one parameter group of analytical diffeomorphisms 
$t\mapsto \varphi^t$ which preserve 
$S$ but move $\gamma(s)$ non-trivially for each $t$ and we just need 
to take $\varphi_k=\varphi^{1/k}$. But this implies that
\be \label{4.70}
[Z^n_t(S,s)]_{\nu\nu'}=[Z^n_t(S,\varphi_k(s))]_{\nu\nu'}
\ee
for all $k=0,1,2,..$. Due to the mutual orthogonality of spin network 
functions over mutually different graphs, (\ref{4.70})
contradicts normalizability (\ref{4.64}) unless $[Z^n_t(S,s)]_{\nu\nu'}=0$
for such $s$ since 
\be \label{4.71}
||M^n_t(S)_{\nu\nu'}||^2
\ge \sum_{s',\gamma(s')=\gamma(s)}|[Z^n_t(S,s')]_{\nu\nu'}|^2
\ge \sum_{k=0}^\infty |[Z^n_t(S,\varphi_k(s))]_{\nu\nu'}|^2
= |[Z^n_t(S,s)]_{\nu\nu'}|^2 \sum_{k=0}^\infty\;1
\ee

We conclude that $M^n_t(S)$ is a matrix  of cylindrical $L_2-$functions 
over the graph $\partial S$ in $D=3$ or over $S$ in $D=2$ and it is 
a constant function in $D>3$. Hence we may write it in the form 
\be \label{4.72}
M^n_t(S)_{\nu\nu'}=K^n_t(S)_{\nu\nu'}+\sum_I K^n_t(S,I)_{\nu\nu'}
T_{\gamma_S,I}
\ee
where $K^n_t(S)_{\nu,\nu'},\;K^n_t(S,I)_{\nu,\nu'}$ are constants,
$\gamma_S=\overline{S}$ in $D=2$, $\gamma_S=\partial S$ in $D=3$, 
$I$ denotes a sum over spin network labels other than the graph
and of course $K^n_t(S,I)_{\nu,\nu'}=0$ for $D>3$. Since 
$\partial\varphi(S)=\varphi(\partial S)$ we find that these constant
matrices only depend on the diffeomorphism type $\hat{S}$ of the surface 
$S$, that is,
\be \label{4.73}
K^n_t(S)_{\nu,\nu'}=K^n_t(\hat{S})_{\nu,\nu'}
\mbox{ and }
K^n_t(S,I)_{\nu,\nu'}=K^n_t(\hat{S},I)_{\nu,\nu'}
\ee

Let us now consider the cases $D=2,3$ more closely. Since by construction
our Weyl algebra of fluxes is built from the fluxes through a disjoint 
union of cubes $\Box$, the associated $\pi(W^n_t(\Box))$ are mutually 
commuting and it will be sufficient to consider each $\Box$ separately.
We may write 
\be \label{4.74a}
[M^n_t(\Box)](A)= \left\{ \begin{array}{ll}
\rho^n_t(A(\overline{\Box})) & D=2\\
\rho^n_t(A(\partial\Box)) & D=3
\end{array}
\right.
\ee
where we have dropped the label $\hat{\Box}$ since the diffeomorphism class
types of all our $\Box$ coincide. Now we can subdivide 
$\Box=\Box_1\cup\Box_2$ into two disjoint pieces. Since the corresponding 
Weyl operators commute and since $[W^n_t(\Box_i),M^n_t(\Box_j)_{\nu\nu'}]=0$
because the edges of the graph $\bar{\Box}_j$ or $\partial\Box_j$ are 
either of the ``in'' or ``out'' type with respect to $\Box_i$, we easily 
find 
\ba \label{4.74}
D=2& :& 
\rho^n_t(A(\bar{\Box}_1))\rho^n_t(A(\bar{\Box}_2))
=\rho^n_t(A(\bar{\Box}))
=\rho^n_t(A(\bar{\Box}_1\circ\bar{\Box}_2))
=\rho^n_t(A(\bar{\Box}_1)A(\circ\bar{\Box}_2))
\nonumber\\
D=3 &: & 
\rho^n_t(A(\partial\Box_1))\rho^n_t(A(\partial\Box_2))
=\rho^n_t(A(\partial\Box))
=\rho^n_t(A(\partial\Box_1\circ\partial\Box_2))
=\rho^n_t(A(\partial\Box_1)A(\partial\Box_2))
\ea
where we have chosen appropriate starting points of the edges 
$\bar{\Box}$ and loops $\partial\Box$ respectively. 

Since $A\in\ab$ is 
arbitrary we find that for arbitrary $h_1,h_2\in G$
\be \label{4.75}
\rho^n_t(h_1)\rho^n_t(h_2)=
\rho^n_t(h_2)\rho^n_t(h_1)=\rho^n_t(h_1 h_2)
\ee
where commutativity follows from the commutativity of the corresponding
Weyl operators. Setting e.g. $h_1=1_G$ we see that 
$\rho^n_t(1_G)=\pi(1)$ is the identity operator. It follows that 
$h\mapsto \rho^n_t(h)$ is a commutative representation of $G$ on the 
not necessarily separable Hilbert space $\ell_2({\cal N})$ where $\cal N$ 
denotes the 
countable index set of the labels $\nu$. Let us denote the inner product
on $l_2({\cal N})$ by $(.,.)'$, that is 
$(v,v')'=\sum_\nu \bar{v}_\nu v'_\nu$. Consider the new inner product
\be \label{4.75a}
(v,v'):=\int_G d\mu_H(g) (\rho^n_t(g)v,\rho^n_t(g)v')'
\ee
We must check whether (\ref{4.75a}) is well defined.
The vectors $1^\nu$ form a complete orthonormal basis in $l_2$ with 
respect to the inner product $(.,.)'$ and the new Hilbert space is the 
completion with respect to the new inner product of the finite linear 
combinations of the $1^\nu$, so it 
suffices to check that $(1^\nu,1^\nu)<\infty$ for all $\nu$. We have 
\be \label{4.75b}
(1^\nu,1^\nu)=\int_G d\mu_H(g) \sum_{\nu'} 
|(\rho^n_t)_{\nu'\nu}(g)|^2 
\le\sum_{\nu'} ||M^n_t(\hat{\Box})_{\nu'\nu}||^2_{\mu_0}=1
\ee
by (\ref{4.64}) because $M^n_t(S)$ depends on the connection 
only through the edge $\bar{\Box}$ or the loop $\partial\Box$ respectively.
Hence (\ref{4.75a}) is well-defined and $g\mapsto \rho^n_t(g)$ is a unitary
representation of $G$ on this Hilbert space with inner product $(.,.)$.

Since $G$ is compact, $\rho^n_t$, represented on that Hilbert space is 
unitarily equivalent 
to a (possibly uncountably) direct sum of irreducible, finite dimensional 
representations \cite{23a} (proposition 2.5 and theorem 3.1) all of 
which must be commutative. 
If $G$ is not Abelian, then the only commutative irreducible 
representations are trivial and it follows immediately 
$\rho^n_t(h)=\pi(1)$ for all $h\in G$. If $G$ is Abelian then 
$G=U(1)^N$ for some $N$ and every irreducible representation is of the 
form $(u_1,..,u_N)\mapsto (u_1^{z_1},..,u_N^{z_N})$ for some integers
$z_k$ and any $u_k\in U(1)$. In our case the representation of every $U(1)$ 
factor 
that occurs in the decomposition of $\rho^n_t(h)$ into irreducibles is 
therefore of the form $u\mapsto u^{z^{n}_t}$ where $z^n_t\in \Zl$ and 
$u\in U(1)$. Due to the representation property 
$\pi(W^n_s(\Box))\pi(W^n_t(\Box))=\pi(W^n_{s+t}(\Box))$ for all 
$s,t\in\Rl$ and due to the fact that all edges in question are of the 
``in'' or ``out'' type with respect to $S$ we infer that 
$\rho^n_s(h)\rho^n_t(h)=\rho^n_{s+t}(h)$ is a one-parameter group of 
representations. This implies that $z^n_{s+t}=z^n_s+z^n_t$ for any 
$s,t\in \Rl$. Due to weak continuity we have $\rho^n_t(h)\to \pi(1)$ 
as $t\to 0$. Since $z^n_t$ is an integer, there exists $\epsilon^n>0$ 
such that $z^n_t=0$ for all $|t|<\epsilon^n$. But then for any $t\in 
\Rl$ we find $m\in \Nl$ such that $|t/m|<\epsilon^n$ and thus 
$z^n_t=m\;z^n_{t/m}=0$. Thus, also in the Abelian case the only 
occurring representation is trivial and we also get here that 
$\rho^n_t(h)=\pi(1)$.

It remains to discuss the case $D>3$. Since in this case 
$M^n_t(\Box)=M^n_t(\hat{\Box})$ is just a constant we have by 
splitting $\Box=\Box_1\cup\Box_2$ into disjoint pieces that 
\be \label{4.76}
M^n_t(\hat{\Box})=M^n_t(\Box)=M^n_t(\Box_1\cup\Box_2)
=M^n_t(\Box_1)\;M^n_t(\Box_2)
=M^n_t(\hat{\Box})^2
\ee
and since $M^n_t(\hat{\Box})$ is invertible we find 
$M^n_t(\Box)=\pi(1)$.

We conclude that $M^n_t(S)=\pi(1)$ for $n^j(x)=n^j=const.$ and any (allowed)
surface $S$.
For an arbitrary unit vector $n_j(x)$ we find a constant unit vector 
$n^0_j$ and an element $g_{n,n^0}\in$Fun$(\Man,G)$ such that 
$\alpha_{g_{n,n^0}}(W^n_t(S))=W^{n^0}_t(S)$. Then 
\ba \label{4.77}
\pi(W^n_t(S))
&=& \pi(\alpha_{g_{n,n^0}^{-1}}(W^{n_0}_t(S)))
=U_\pi(g_{n,n^0})^{-1}\pi(W^{n_0}_t(S))U_\pi(g_{n,n^0})
\\
&=& 
U_\pi(g_{n,n^0})^{-1}[W^{n_0}_t(S)\otimes\pi(1)] U_\pi(g_{n,n^0})
=[\alpha_{g_{n,n^0}^{-1}}(W^{n_0}_t(S))\otimes\pi(1)] 
=[W^n_t(S)\otimes\pi(1)] 
\nonumber
\ea
so that $M^n_t(S)=\pi(1)$ also in the general case.\\
\\
We thus have shown that $\pi(W^n_t(S))=W^n_t(S)\otimes\pi(1)$. We can now 
finally invoke irreducibility: If the representation is to be irreducible,
then every vector is cyclic, in particular any of the $1^\nu$ is cyclic.
But the algebra of operators generated by $\pi(f),\pi(W^n_t(S))$ never 
leaves the sector ${\cal H}_\nu={\cal H}_0\otimes 1^\nu$. It follows that 
we can allow only one copy of the Ashtekar-Lewandowski Hilbert space.
That ${\cal H}_0$ itself is the representation space of an irreducible 
representation of $\mathfrak{A}$ will be shown in \cite{24}.\\
\\
This finishes the proof.\\
$\Box$\\
\\
\\
\\
{\large Acknowledgments}\\
\\
We thank the Erwin -- Schr\"odinger Institute
for providing a friendly hospitality and a challenging research 
environment during the 'Quantum field theory on curved spacetimes'
workshop, where part of this work was completed. 
Thanks also go to the organizers of this workshop, Klaus
Fredenhagen, Robert Wald, and Jacob Yngvasson, as well as to
the other guests, especially Abhay Ashtekar, Klaus
Fredenhagen, and Jurek Lewandowski, for stimulating discussions
concerning the questions addressed in this work. 
H.S. also thanks the Studienstiftung des Deutschen Volkes for financial
support, and the Albert-Einstein Institut for hospitality.
This work was supported in part by the NSF grant PHY-0090091.   

\begin{appendix}

\section{A simpler proof under stronger assumptions}
\label{sa}

In this appendix we present a somewhat simpler proof of theorem 
\ref{th4.2} which works under the following assumption in addition to 
i), ii), iii) and iv) specified there:\\
\\
v) {\bf Gauge Invariance}\\
The representation $U_\pi$ of Fun$(\Sigma,G)$ is unitary and coincides 
with the natural representation.\\
\\
Under this assumption we can avoid step 2 in the proof of the theorem 
altogether by means of the following observation:\\
\\
While the proof in the main text never uses gauge invariance, 
the following replacement of the non-Abelean part of step 3 
heavily relies on gauge invariance of the $1^\nu$ which follows from 
our assumption v) made above.

According to assumption iv) we may let $\phi(t)\to \delta(t,0)$ 
in (\ref{4.41}) and (\ref{4.42}) and in 
this limit the $(S,n)-$dependence of all the states 
$1^\nu_{\phi,S_n}\to 1^\nu$ vanishes.

Let $\varphi^{S,e}_m$ be an analytic
diffeomorphism that makes $\alpha_{\varphi^{S,e}_m}(T_s)$ intersect with
$S$ in precisely $m$ interior points $p_k$ of $\varphi^{S,e}_m(e)$, say
with positive relative orientation. Then (\ref{4.41}) implies
\ba \label{4.58}
&&
\sum_j
<1^\nu,\pi(Y(S_j)^2 \alpha_{\varphi^{e,S}_m}^{-1}(T_s)) 1^{\nu'}>
\nonumber\\
&=&8
<1^\nu,\pi(-m\lambda_{\pi_e} T_s
+\sum_j \sum_{I<J}^m R^n_{e_I} R^n_{e_J}T_s)
1^{\nu'}>
\nonumber\\
&=&
\sum_j
[<1^\nu,\pi(\alpha_{\varphi^{S,e}_m}^{-1}(T_s))
\pi(Y_j(S)^2)1^{\nu'}>
+2<\pi(Y_j(S))1^\nu,\pi(
\alpha_{\varphi^{S,e}_m}^{-1}(T_s))\pi(Y_j(S))1^{\nu'}>
\nonumber\\
&& +<\pi(Y_j(S)^2)1^\nu,\pi(
\alpha_{\varphi^{S,e}_m}^{-1}(T_s))1^{\nu'}>]
\ea
Let $G_p:=$Fun$(\{p_k\}_{k=1}^m,\sigma)$ be the subgroup of the gauge
group 
which is trivial except at the points $p_k$. Due to the gauge invariance
of the $1^\nu$ and the normalization of the Haar measure we have for the
left hand side of (\ref{4.58})
\ba \label{4.59}
&& 8
<1^\nu,\pi(-m\lambda_{\pi_e} T_s
+\sum_j \sum_{I<J}^m R^j_{e_I} R^j_{e_J}T_s)1^{\nu'}>
\nonumber\\
&=& 8 \int_{G_p} d\mu_0(g)
<U_\pi(g)1^\nu,\pi(-m\lambda_{\pi_e} T_s
+\sum_j \sum_{I<J}^m R^j_{e_I} R^j_{e_J}T_s)U_\pi(g)1^{\nu'}>
\nonumber\\
&=& 8 \int_{G_p} d\mu_0(g)
<1^\nu,\pi(\alpha_{g^{-1}}(-m\lambda_{\pi_e} T_s
+\sum_j \sum_{I<J}^m R^j_{e_I} R^j_{e_J}T_s))1^{\nu'}>
\nonumber\\
&=&
8\int_{G^m} d\mu_H(g_1)..d\mu_H(g_m)
<1^\nu,\pi(-m\lambda_{\pi_e} T_s
+\sum_j \sum_{I<J=}^m
\mbox{ad}_{jk}(g_I)\mbox{ad}_{jl}(g_J)
R^k_{e_I} R^l_{e_J}T_s)1^{\nu'}>
\nonumber\\
&=&
-8m\lambda_{\pi_e} <1^\nu,\pi(T_s)1^{\nu'}>
\ea
where in the third step we have used $G_p-$invariance of $T_s$ and in the
fourth the Peter\&Weyl theorem. Here $\mu_0$ is the induced Haar
measure on $G_p$. Hence, since the right hand side of (\ref{4.58}) is again 
independent of $m$, we conclude $\mu_\nu=\mu_0$.

Thus, averaging over the gauge group has
the same effect as averaging over the discrete subset of the diffeomorphism 
group that was constructed in step 2 which therefore becomes redundant. \\
\\
In conclusion, this way of proving theorem \ref{th4.2} uses 
step 1, avoids step 2 altogether, modifies 
the non-Abelean part of step 3 as just displayed and leaves step 4
unmodified.

\section{Construction of more general $C^\infty-$vectors for the Fluxes}
\label{sb}

In order to supplement the discussion in the main text around the 
non-Abelean part of step 3 in the proof of theorem \ref{th4.2} we list 
here some additional information about the construction of more general
$C^\infty-$vectors for the fluc operators\footnote{In order to use them 
one would need to replace in assumption iv) of theorem \ref{th4.2} 
the $Y_n(S)$ by the $Y^+_n(S)$ constructed below.}.\\
A)\\
If $n_j(x)=const.$ we may also get rid of the $n-$dependence of the 
$\psi_{\phi,S_n}$ in (\ref{4.14}) 
as follows: We notice that each flux vector field 
can be uniquely split as $Y_n(S)=Y^+_n(S)-Y^-_n(S)$ where 
\be \label{4.18a}
Y^\pm_n(S)=
\sum_{x\in S} n_j(x)\sum_{[e]\in {\cal E}^\pm_x(S)} R^j_{x,[e]}
=:\sum_{x\in S} n_j(x) R^j_{x,S,\pm}
\ee
and ${\cal E}^\pm_x(S)=\{[e]\in {\cal E};\;\sigma([S]_x,[e])=\pm 1\}$.
It is easy to check that for the semisimple generators 
\be \label{4.18b}
Y^\pm_j(S)=\frac{1}{2}([Y_k(S),Y_l(S)]f_{klj}\pm Y_j(S)
\ee
where $Y_j(S)=Y_n(S)$ for $n_k=\delta_{jk}$ so that the algebra generated 
by the $Y_n(S)$ allows us to isolate the pieces $Y^\pm_j(S)$.
For the Abelian generators we define $Y^+_j(S)=Y_j(S)$. The $Y^+_j(S)$
satisfy the simple algebra
\be \label{4.18c}
[Y^+_j(S),Y^+_k(S)]=f_{jkl} Y^+_l(S)
\ee
(Actually, since Lie$(S)$ is semisimple, any $A\in$Lie$(S)$ can be written 
as $A=[B,C]$ so that we can also isolate the $Y^\pm_n(S)$ for any $n$ 
such that $n^j\tau_j\in \mbox{Lie}(S)$.) 
Since Vec contains them, we are allowed to construct the corresponding 
Weyl elements
\be \label{4.18d}
W_t(S):=\exp(t^j Y^+_j(S))
\ee
and since the $Y^+_j(S)$ have the same commutation relations as the 
Lie algebra basis elements $\tau_j$ we conclude 
\be \label{4.18e}
W_t(S) W_{t'}(S):=W_{c(t,t')}(S)
\ee
where the composition function $c$ is defined as follows:
For any compact, connected gauge group the exponential map 
$t^j\mapsto g(t):=\exp(t^j\tau_j)$
is surjective,
hence we find a subset $R\subset \Rl^{|\dim(G)|}$ so that it becomes 
a bijection. We then may define $c:\;R\times R\to R$ uniquely by 
\be \label{4.18f}
e^{c^j(t,t')\tau_j}:=e^{t^j\tau_j} \;e^{t^{\prime j}\tau_j}
\ee
Then (\ref{4.18e}) follows.
Consider now for any $\psi\in {\cal H}$ and $\phi\in C^\infty(G)$ the vector
\be \label{4.18g}
\psi_{\phi,S}:=\int_G d\mu_H(g) \phi(g) \pi(W_g(S))\psi
\ee
where $W_{g(t)}(S):=W_t(S)$. Then (\ref{4.18a}) is a $C^\infty$ vector for 
all $\dim(G)$ operators $\pi(Y^+_j(S))$, namely
\be \label{4.18h}
-i\pi(Y^+_j(S))\psi_{\phi,S}=\psi_{-R^j\phi,S}
\ee
due to the invariance of the Haar measure. It follows by similar arguments 
as those displayed in corollary \ref{col4.1} that the vectors 
$\pi(f)1^\nu_{\phi,S}$ provide a common 
dense domain ${\cal D}(S)$ of $C^\infty-$vectors for all $Y^+_j(S)$.\\
\\
B)\\
By methods similar to those displayed in equations (\ref{4.41}) and 
(\ref{4.42}) one can show that symmetry of the fluxes implies
for $n^j(x)=const.$
\be \label{4.43}
-<1^\nu_{\phi,S},\pi(Y^+_n(S)f') 1^{\nu'}_{\phi',S}>
= <1^\nu_{\phi,S},\pi(f') 1^{\nu'}_{-R_n\phi',S}>
+ <1^\nu_{-R_n\phi,S},\pi(f') 1^{\nu'}_{\phi',S}>
\ee
and
\ba \label{4.50}
&&<1^\nu_{\phi,S},\pi((Y^+_n(S))^2f) 1^{\nu'}_{\phi',S}>
\nonumber\\
&=& <1^\nu_{\phi,S},\pi(f) 1^{\nu'}_{(R_n)^2\phi',S}>
+ 2<1^\nu_{-R_n\phi,S},\pi(f) 1^{\nu'}_{-R_n \phi',S}>
+ <1^\nu_{R_n^2\phi,S},\pi(f) 1^{\nu'}_{\phi',S}>
\ea
where now the $n-$independent $C^\infty-$vectors of A) are used.

\end{appendix}

\end{document}